\documentclass[preprint,showpacs,preprintnumbers,amsmath,amssymb]{revtex4}
\usepackage{latexsym,amsmath,tabularx,amssymb,bm}
\usepackage{epsfig}
\usepackage{citesort}

\long\def\comment#1{ }

\usepackage{graphicx}
\usepackage{amssymb}
\usepackage{graphicx}

\def\simge{\mathrel{%
    \rlap{\raise 0.511ex \hbox{$>$}}{\lower 0.511ex \hbox{$\sim$}}}}
\def\simle{\mathrel{
    \rlap{\raise 0.511ex \hbox{$<$}}{\lower 0.511ex \hbox{$\sim$}}}}

\newcommand \beq{\begin{equation}}
\newcommand \eeq{\end{equation}}
\newcommand{\del}{\partial}

\begin{document}

\title{On the 2-point function of the $O(N)$ model}

\author{Federico Benitez}
\email{federico@fisica.edu.uy}
\affiliation{Instituto de F\'isica, Facultad de Ciencias, Universidad de la Rep\'ublica,
Igu\'a 4225, 11000, Montevideo, Uruguay}
\author{Ram\'on M\'endez Galain}
\email{mendezg@fing.edu.uy}
\affiliation{Instituto de F\'{\i}sica, Facultad de Ingenier\'{\i}a, Univ.~de la Rep\'ublica, J.H.y Reissig 565, 11000 Montevideo, Uruguay}
\author{Nicol\'as Wschebor}
\email{nicws@fing.edu.uy}
\affiliation{Instituto de F\'{\i}sica, Facultad de Ingenier\'{\i}a, Univ.~de la Rep\'ublica, J.H.y Reissig 565, 11000 Montevideo, Uruguay}

\date{\today}
\vspace{0.8cm}
\begin{abstract}
The self-energy of the critical 3-dimensional $O(N)$ model is
calculated. The analysis is performed in the context of the
Non-Perturbative Renormalization Group, by exploiting an approximation which takes into account contributions of an
infinite number of vertices. A very simple calculation
yields the $2$-point function in the whole range of momenta, from the UV
Gaussian regime to the  scaling one. Results are
in good agreement with best estimates in the literature for any
value of $N$ in all momenta regimes. This encourages the use of this simple
approximation procedure to calculate correlation functions at finite momenta in other physical
situations.
\end{abstract}

\maketitle
\newpage

\section{Introduction}
\label{introduction} The $O(N)$ scalar model describes many
phenomena in a wide range of physical situations. Besides the $N=1$
case, which corresponds to Ising like systems, with a wide range of
applications as, e.g., liquid-gas transition, the $N=2$ model
describes superfluid Helium, $N=3$ can be used to study
ferromagnets, $N=4$ allows the study of the Higgs sector of the
Standard Model at finite temperature, and the $N=0$ case describes
the physics of some polymers \cite{Guida98}. As a natural
consequence, a huge amount of work has been devoted to the study of
this family of models. Leaving aside the $d=2$ case, where specific
methods exist, most of the existing results correspond to
thermodynamical  properties, as critical exponents or phase
diagrams, i.e., physical quantities encoded in correlators at small
external momenta, e.g., the effective potential. With this goal,
very complicated techniques, as resummed perturbative calculations
carried up to 7-th order \cite{Guida98,Butti04}, high--temperature
expansions  \cite{Campostrini01,Campostrini02,Campostrini06}, or
Monte-Carlo methods
\cite{Hasenbusch01,Campostrini01,Deng03,Hasenbusch05,Campostrini06},
were used. When instead trying to get physical quantities depending
on finite momenta, such as the self-energy of the model, fewer
results can be found in the literature (see for example
\cite{Pelissetto02} and references therein). All these calculations
suffer from a common difficulty, which is general to a vast class of
problems: it is extremely nontrivial to deal with systems having
highly correlated components. In this sense, as the $O(N)$ model is
simpler than most other such problems, it has been largely used as a
testing ground for the development of calculation schemes in
non-perturbative contexts.

The Non-Perturbative Renormalization Group (NPRG)
\cite{Wetterich92,Ellwanger93,Tetradis94,Morris94b,Morris94c} is a
general framework conceived to deal with this kind of situations.
It is based in an infinite set of exact equations giving all
renormalized correlation functions between the various components of
a given system. Naturally, as one has to deal with an infinite tower
of coupled differential equations, in order to solve them the use of
approximations is unavoidable. Several years ago, a systematic
approximation scheme was developed
\cite{Morris94c,Berges00,Bagnuls00} which allows for the solution of
this set of equations in a particular case: The so called derivative
expansion (DE) is based in an expansion in the powers of the
derivatives of the fields. Even if there is no formal proof of its
convergence, the DE has provided very competitive results in
problems where only small (eventually zero) external momenta play a
role. Among many other applications (see, e.g.,
\cite{Berges00,Delamotte03,Delamotte04}), the approximation was
applied to the $O(N)$ model
\cite{Gollisch:2001ff,Litim,VonGersdorff00,Berges00,Berges95,Morris97},
even up to the next-next-to-leading order of the scheme in the $N=1$
Ising case \cite{Canet03}, yielding at this order critical exponents
of a similar quality as those obtained using 7-loops resummed
perturbative calculations.

On the other hand, when trying to describe phenomena involving all
modes, the situation is different. For example, to get the
transition temperature of a dilute gas to a Bose-Einstein
condensate, one needs the self-energy of the $O(2)$ model in
3-dimensions at arbitrary momenta \cite{Baym99}; relevant
information comes from the intermediate momentum region between the
IR and the UV ranges. Within the NPRG, only calculations including a
finite number of vertices \cite{weinberg73} had been considered up
to now, either in $O(N)$
\cite{Ledowski04,kopietz,Blaizot04,Blaizot05,Blaizot06} or in more
involved problems such as QCD \cite{Fischer04,truncation}.

Recently, a general approximation scheme suitable to get any $n$-point
function at any finite momenta within the NPRG has been proposed
\cite{BMW}. The strategy has many interesting similarities with DE.
First, it can be applied in principle to any model. Second, although
the approximation is not controlled by a small parameter, it can be
systematically improved. Furthermore, this strategy reproduces both
perturbative and DE results, in their corresponding limits; for
example, if solving the $2$-point function flow equation at the
leading order (LO) of the procedure, the $2$-point function includes
all $1$-loop contributions while the effective potential includes
all $2$-loops ones. It is possible to apply, on top of the approximation presented in
\cite{BMW}, an expansion in powers of the field, as frequently done
in DE: at least in the studied case, this expansion seems to
converge rapidly \cite{Guerra07}. Nevertheless, when considering
only the first order of the expansion, the correct result for a
quantity such as the critical exponent $\eta$ can be missed by as
much as 60\%. Thus, as an expansion in powers of the field
corresponds to an expansion in the number of vertices
\cite{Guerra07}, the latter remark is a strong support to
approximations, as that of \cite{BMW}, which simultaneously include
an infinite number of vertices.

The LO of the procedure was used in \cite{BMWnum} in order to
calculate the $2$-point function of the $N=1$ case. Following a
simple (and yet accurate) strategy, it was shown that, within an
analytical and numerical effort similar to that of the DE, one gets
a self-energy with the correct shape at all momentum regimes: One
gets the logarithmic UV behavior; its pre-coefficient, which is in
fact a $2$-loop quantity, follows with only 8\% error. In the IR,
critical exponents are obtained with a quality similar to that of DE
at NLO. As for the intermediate crossover regime, a quantity
sensitive to this range of momenta was calculated to get a result
close to the error bars of both lattice and resummed $7$-loop
calculations.

The purpose of this paper is to apply the method presented in
\cite{BMW}, at its LO, to the $O(N)$ model. In \cite{BMW} it was
shown that the LO of the procedure is already exact in the large $N$
limit of the model. Moreover, in this limit, a simple analytical
solution of the $n$-point functions at finite momenta was presented.
Here, we shall implement the method, at any value of $N$, in order
to numerically solve the approximate flow equations of the $2$-point
function, at criticality and in $d=3$. We shall
follow a simple strategy, similar as that used in \cite{BMWnum}.

The paper is organized as follows. In the next section we shall
present the general approximation procedure introduced in \cite{BMW}
in the framework of a field theory with $N$ boson fields and, in
particular, when the model has $O(N)$ symmetry. In section
\ref{strategies}, we shall present two possible strategies to solve
the $2$-point function equations: the first one is simpler, but it
looses the above mentioned 2-loop exactness of the effective
potential; in the second strategy, with a slight increase in the
numerical effort, the 2-loop exactness is recovered. In section
\ref{results}, we present our results, both in the scaling sector
and at large and intermediate momenta; in particular, we calculate
some quantities to gauge the quality of the $2$-point function thus
obtained, and compare our results with those following from other
means. Finally, in section \ref{largeN}, we study analytically the
large $N$ behavior of our results, both at leading and
next-to-leading order in $1/N$, and compare them with numerical
results of the previous section, as well as with exact results known
in the literature.

\section{The Approximation Scheme}

\label{appscheme}

In this section we shall briefly present the general formalism of
the NPRG and describe the approximation scheme to calculate
$n$-point functions at finite momenta introduced in \cite{BMW}. We
shall make the presentation considering in the first place a generic
Euclidean field theory with $N$ boson fields $\varphi_i$, denoted
collectively by $\varphi$, with action $S[\varphi]$. Then, we shall
specialize to the case where $S[\varphi]$ has an $O(N)$ symmetry.

The NPRG equations relate the bare action to the full effective
action. This relation is obtained by controlling the magnitude of
long wavelength field fluctuations with the help of an infrared
cut-off, which is implemented
\cite{Tetradis94,Ellwanger94a,Morris94b,Morris94c} by adding to the
bare action $S[\varphi]$ a regulator of the form
\begin{equation}
  \Delta S_\kappa[\varphi] =\frac{1}{2} \int \frac{{\rm d}^dq}{(2\pi)^d}
(R_\kappa)_{ij}(q)
\varphi_i(q)\varphi_j(-q),
\end{equation}\normalsize
where $(R_\kappa)_{ij}(q)$ denotes a family of ``cut-off functions''
depending on a parameter $\kappa$; above, a sum over repeated indices is understood.  The role of $\Delta S_\kappa$
is to suppress the fluctuations with momenta $q\simle \kappa$,
while leaving unaffected the modes with $q\simge \kappa$. Thus,
typically $ (R_{\kappa})_{ij}(q) \sim \kappa^2 \delta_{ij}$ when $ q \ll \kappa$, and
$(R_{\kappa})_{ij}(q)\to 0$
 when $ q\simge \kappa$.

One can define an effective average action corresponding to
$S[\varphi]+\Delta S_\kappa[\varphi]$  by $\Gamma_\kappa[\phi]$,
where $\phi$ is the average field in presence of external sources,
$\phi_i(x)=\left\langle \varphi_i(x) \right\rangle$. When
$\kappa=\Lambda$, with $\Lambda$ a scale much larger than all other
scales in the problem, fluctuations are suppressed and
$\Gamma_\Lambda[\phi]$ coincides with the classical action. As
$\kappa$ decreases,
 more and more fluctuations are taken into account and, as  $\kappa\to 0$, $\Gamma_{\kappa=0}[\phi]$ becomes the usual effective action $\Gamma[\phi]$ (see e.g. \cite{Berges00}). The variation
with $\kappa$ of $\Gamma_\kappa[\phi]$ is governed by the
following flow
equation \cite{Tetradis94,Ellwanger94a,Morris94b,Morris94c}:
\begin{equation}
\label{NPRGeq}
\partial_\kappa \Gamma_\kappa[\phi]=\frac{1}{2}\int \frac{d^dq}{(2\pi)^d} \mathrm{tr} \bigg\{ \partial_\kappa R_\kappa(q^2)
\left[\Gamma_\kappa^{(2)}+R_\kappa\right]^{-1}_{q;-q}\bigg\},
\end{equation}
where $\Gamma_\kappa^{(2)}$ denotes the matrix of second derivatives of $\Gamma_\kappa$ w.r.t. $\phi$ (i.e., the matrix of components ${(\Gamma_\kappa^{(2)})}_{ij}= \delta^2\Gamma_\kappa/\delta \phi_i \delta \phi_j$) and
the trace is taken over internal indices.

For a given value of $\kappa$, we define the $n$-point vertices $\Gamma_\kappa^{(n)}$ in a constant external
field $\phi$:
\begin{multline}
(2\pi)^d \delta^{(d)}\Big(\sum_j p_j\Big){(\Gamma^{(n)}_\kappa)}_{i_1,i_2,\dots,i_n}(p_1,\dots,p_n;\phi)
\\ =\int d^dx_1\dots\int d^dx_n
e^{i\sum_{j=1}^n p_jx_j}\frac{\delta^n\Gamma_\kappa}{\delta\phi_{i_1}(x_1)
\dots \delta\phi_{i_n}(x_n)} \bigr\vert_{\phi(x) \equiv \phi}.
\end{multline}

\begin{figure}[t]
\begin{center}
\includegraphics*[scale=0.7]{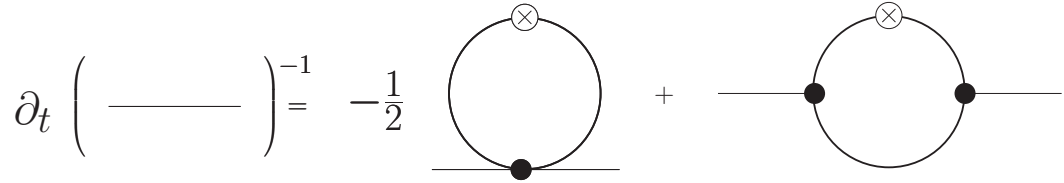}
\end{center}
\caption{ \label{diagrams} A diagrammatic representation of the flow
equation for the two-point function in an external field. The lines
and dots represent full propagators and vertices. Crosses represent
the insertion of $\partial_t R_\kappa$.}
\end{figure}

By differentiating eq.~(\ref{NPRGeq}) with respect to
$\phi_{i_1}(x_1), \cdots , \phi_{i_n}(x_n)$ and then letting the
field be constant, one gets the flow equations for all $n$-point
functions in a constant background field $\phi$.  These equations
can be represented diagrammatically by  one loop  diagrams with
dressed vertices and propagators (see e.g. \cite{Berges00}). For
instance, the flow of the 2-point function in a constant external
field reads:
\begin{eqnarray}
\label{gamma2}
\partial_t \Gamma_{ab}^{(2)}(p;\phi)&=&\int
\frac{d^dq}{(2\pi)^d} \partial_t (R_\kappa)_{in}(q)\left\{G_{ij}(q;\phi)\Gamma_{ajk}^{(3)}(p,q,-p-q;\phi)\right. \nonumber \\
&&\times G_{kl}(q+p;\phi)\Gamma_{blm}^{(3)}(-p,p+q,-q;\phi)G_{mn}(q;\phi) \nonumber \\
&&\left.-\frac{1}{2}G_{ij}(q;\phi)\Gamma_{abjk}^{(4)}(p,-p,q,-q;\phi)G_{kn}(q;\phi)\right\}
\end{eqnarray}
where $G$ is the matrix of propagators:
\begin{equation}\label{G-gamma2}
G^{-1}_\kappa (q^2;\phi) = \Gamma^{(2)}_\kappa (q,-q;\phi) + R_\kappa(q^2) .
\end{equation}
The diagrammatic representation of eq.~(\ref{gamma2}) is given in
figure \ref{diagrams}. Above, we have introduced the dimensionless
variable $t\equiv \ln (\kappa/\Lambda)$. From now on, as we already
did in eq. (\ref{gamma2}), the $\kappa$ dependence of the $n$-point
functions shall not be made explicit, unless necessary to avoid
confusions.

Flow equations for the $n$-point functions do not close: for example, in order to solve eq.~(\ref{gamma2}) one needs
the 3- and the 4-point functions, $\Gamma_\kappa^{(3)}$ and $\Gamma_\kappa^{(4)}$ respectively. However, in ref. \cite{BMW} an approximation scheme was introduced in order to solve flow equations yielding $n$-point functions at finite momenta. In doing so,
it is possible to exploit
two properties of these flow equations:  i)  due to the factor $\partial_t R_\kappa(q)$ in the loop integral, the integration
is dominated by momenta $q\simle \kappa$; ii)  as they are regulated in the IR, $n$-point vertices are smooth functions of momenta. These two properties allow one to make an expansion in powers of $q^2/\kappa^2$, independently of the value  of the external momenta $p$. As a typical $n$-point function entering the flow has the form $\Gamma^{(n)}_\kappa(p_1,p_2,...,p_{n-1}+q,p_n-q;\phi)$, where $q$ is the loop momentum, then the \emph{leading order} (LO) of the approximation scheme consists in neglecting the  $q$-dependence of such vertex functions:
\begin{equation}\label{approx}
\Gamma^{(n)}(p_1,p_2,...,p_{n-1}+q,p_n-q;\phi)\sim
\Gamma^{(n)}(p_1,p_2,...,p_{n-1},p_n;\phi).
\end{equation}
Note that this approximation is a priori  well justified.
Indeed, when all the external momenta $p_i$ are zero, this kind of approximation
is at the basis of DE which, as discussed above, turns out to be a good
approximation. When the external momenta $p_i$ start to grow, the
approximation in eq.~(\ref{approx}) becomes better and better,  and it is
trivial when all momenta are much larger than $\kappa$. With this
approximation,  eq.~(\ref{gamma2}), for instance, becomes:
\begin{eqnarray}
\label{gamma2app}
\partial_t\Gamma_{ab}^{(2)}(p;\phi)&=&\int
\frac{d^dq}{(2\pi)^d}\partial_t (R_\kappa)_{in}(q)\left\{G_{ij}(q;\phi)\Gamma_{ajk}^{(3)}(p,0,-p;\phi)\right. \nonumber \\
&&\times G_{kl}(q+p;\phi)\Gamma_{blm}^{(3)}(-p,p,0;\phi)G_{mn}(q;\phi) \nonumber \\
&&\left.-\frac{1}{2}G_{ij}(q;\phi)\Gamma_{abjk}^{(4)}(p,-p,0,0;\phi)G_{kn}(q;\phi)\right\}
\end{eqnarray}

Notice that it is not convenient to also assume $q=0$ in the propagators; if this were done, the exactness of the LO of the approximation scheme both at one-loop or large $N$ limit would be lost (see ref. \cite{BMW}).

Now, one can exploit the fact that
\begin{equation}
\label{faireq=0}
\Gamma_{i_1,i_2,\dots,i_n,i_{n+1}}^{(n+1)}(p_1,p_2,...,p_n,0;\phi)=\frac{\partial
\Gamma_{i_1,i_2,\dots,i_n}^{(n)}(p_1,p_2,...p_n;\phi)} {\partial \phi_{i_n+1}}.
\end{equation}
in order to transform eq.~(\ref{gamma2app}) into a {\it closed equation} (recall
that $G_\kappa$ and $\Gamma_\kappa^{(2)}$ are related by
eq.~(\ref{G-gamma2})):
\begin{eqnarray}
\label{2pointclosed}
\partial_t\Gamma_{ab}^{(2)}(p;\phi)&=&\int
\frac{d^dq}{(2\pi)^d} \partial_t (R_\kappa)_{in}(q)\left\{G_{ij}(q,\phi)
\frac{\partial \Gamma_{ak}^{(2)}(p,-p;\phi)} {\partial \phi_{j}}\right. \nonumber \\
&&\times G_{kl}(q+p,\phi)\frac{\partial
\Gamma_{bl}^{(2)}(p,-p;\phi)} {\partial \phi_{m}} G_{mn}(q,\phi)
\nonumber \\
&&\left.-\frac{1}{2}G_{ij}(q,\phi)
\frac{\partial^2
\Gamma_{ab}^{(2)}(p,-p;\phi)} {\partial \phi_j \partial \phi_k}G_{kn}(q,\phi)\right\} .
\end{eqnarray}
Eq. (\ref{2pointclosed}) is valid for an arbitrary theory with $N$ bosonic fields. In the general case, it corresponds to a system of $N(N+1)/2$ equations. From now on, we shall specialize in the particular case where the bare action $S[\varphi]$ has $O(N)$ symmetry. If one chooses $S[\varphi]$ to be renormalizable, it is given by
\beq\label{Sclassical} S[\varphi] = \int {\rm d}^{d}x\,\left\lbrace{
\frac{1}{2}} \left(\del_\mu \varphi_i(x) \del_\mu
\varphi_i(x) \right) + \frac{r}{2} \,  \varphi_i(x) \varphi_i(x)+
\frac{u}{4!} \,\left( \varphi_i(x) \varphi_i(x) \right)^2 \right\rbrace
\,.
\eeq
In order to preserve the $O(N)$ symmetry all along the flow, it is mandatory to consider a regulator respecting the symmetry. Doing so, within this approximation scheme, Ward identities shall be respected throughout all the flow and, in particular, they shall de valid in the ($\kappa \to 0$) physical limit.
The only way to implement this is to consider a diagonal regulator. From now on,

\[
(R_\kappa)_{ij}(q) = R_\kappa(q) \delta_{ij} .
\]

Now, due to the symmetry, the $2$-point
matrix function can be written in terms of only two independent scalar functions; a convenient way to do so is:

\begin{equation}\label{gamma2desc}
\Gamma_{ab}^{(2)} (p, -p; \phi; \kappa) = \Gamma_{A} (p; \rho; \kappa)\delta_{ab} + \phi_a \phi_b \Gamma_B (p; \rho; \kappa)
\end{equation}

\noindent where $\rho(x) = \phi^a(x) \phi^a(x)/2$, $a = 1, \ldots,
N$. Following eq. (\ref{G-gamma2}), a similar decomposition can be done for the propagator matrix. Nevertheless, in this case it proves more convenient to use a decomposition in longitudinal and transverse components
with respect to the external field:
\begin{equation}
G_{ab}(p;\phi;\kappa)=G_T(p;\rho;\kappa)\left(\delta_{ab}-\frac{\phi_a\phi_b}{2\rho}\right)
+G_L(p;\rho;\kappa)\frac{\phi_a\phi_b}{2\rho}.
\end{equation}
It is easy to show that
\begin{align}
G_T^{-1}(p;\rho;\kappa) = &
\Gamma_A(p;\rho;\kappa) + R_\kappa(p), \label{propGTgamma} \\
G_L^{-1}(p;\rho;\kappa) = &\Gamma_A(p;\rho;\kappa)
+ 2\rho \Gamma_B(p;\rho;\kappa)+ R_\kappa(p), \label{propGLgamma}.
\end{align}

Using the definition of the functions
$\Gamma_A$ and $\Gamma_B$, eq. (\ref{gamma2desc}), as well as that of $G_T$ and $G_L$ given above, the flow equation (\ref{2pointclosed}) can be decomposed in two equations for $\Gamma_A$  and $\Gamma_B$:

\begin{equation}\label{gamma_a}
\begin{split}
\partial_t \Gamma_A(p; \rho) & = 2\rho
\Gamma_A'^2(p,\rho) J^{(3)}_{d; L T}(p, \rho)
+ 2\rho \Gamma_B^2(p; \rho) J^{(3)}_{d; T L}(p;\rho)  \\
& - \frac{1}{2} \Big( \big( \Gamma_A'(p;\rho) + 2 \rho
\Gamma_A''(p;\rho) \big) I^{(2)}_{d;L L}(\rho)  \\  & + \big( (N-1)
\Gamma_A'(p;\rho) + 2\Gamma_B(p;\rho) \big) I^{(2)}_{d;T T}(\rho)
\Big)
\end{split}
\end{equation}
\begin{equation}\label{gamma_b} \begin{split}
\partial_t \Gamma_B(p,\rho) &= \big( \Gamma_A'(p;\rho) +
2\Gamma_B(p;\rho) + 2\rho \Gamma_B'(p;\rho) \big)^2
J^{(3)}_{d;LL}(p;\rho)
\\ & + (N-1) \Gamma_B^2(p;\rho) J^{(3)}_{d;TT}(p;\rho) -
\Gamma_A'^2(p;\rho) J^{(3)}_{d;LT}(p;\rho)
\\ & - \Gamma_B^2(p;\rho) J^{(3)}_{d;TL}(p;\rho) - \frac{1}{2} \Big( (N-1) \Gamma_B'(p;\rho) I^{(2)}_{d;TT}(\rho)
\\ & +\big( 5\Gamma_B'(p;\rho) + 2\rho
\Gamma_B''(p;\rho) \big) I^{(2)}_{d;LL}(\rho) \Big) \\ & + \Gamma_B(p;\rho) \;\int
\frac{d^d q}{(2\pi)^d} \left\{
\partial_t R_\kappa(q)  \Gamma_B(q;\rho) \big(G_L(q;\rho) + G_T(q;\rho) \big) G_L(q;\rho)
G_T(q;\rho) \right\}
\end{split}\end{equation}
\noindent Above, and from now on, the prime denotes derivative with respect to $\rho$ and, extending definitions already given in \cite{BMWnum}, we have introduced the functions
\begin{align}
I^{(n)}_{d;\alpha \beta}(\rho;\kappa) = & \int \frac{d^d q}{(2\pi)^d}
 \partial_t R_\kappa(q) G_{\alpha}^{n-1}(q;\rho) G_{\beta} (q;\rho)
 \label{defI} \\
 J^{(n)}_{d;\alpha \beta}(p;\rho;\kappa) = & \int \frac{d^d
q}{(2\pi)^d} \partial_t R_\kappa(q) G_{\alpha}^{n-1}(q;\rho) G_{\beta}
(p+q;\rho) \label{defJ}
\end{align}

\noindent with $\alpha$ and $\beta$ standing  for either $L$ or $T$. In eq.
(\ref{gamma_b}), we made use of the identity

\[
\big(G_T^2(q;\rho) - G_L^2(q;\rho)\big) \frac{1}{\rho} = 2
 \Gamma_B(q;\rho) \big( G_L(q;\rho) + G_T(q;\rho) \big) G_L(q;\rho) G_T(q;\rho)
\]

\noindent in order to render the expressions manifestly regular at $\rho = 0$.

Eqs. (\ref{gamma_a}) and (\ref{gamma_b}) constitute a set of coupled
integro-differential equations, with respect to the real
variables $\kappa$, $\rho$ and the modulus of the momentum $p$.

Before turning to the strategy to solve it, we shall first
comment on an apparent inconsistency of this approximation procedure and the way to avoid it \cite{BMWnum}.
To do so, notice that the $n$-point functions at zero external
momenta can all be considered as derivatives of a single function,
the effective potential $V_\kappa(\rho)$. That is, for example,

\begin{equation} \label{zeromomenta}
\Gamma^{(2)}_{ab}
(p=0;\rho)=\frac{\partial^2 V_\kappa(\rho)}{\partial \phi_a \partial
\phi_b}
\end{equation}
\noindent which entails
\begin{equation}\label{def_gammas}
\Gamma_A(p=0;\rho) = \frac{\partial V_\kappa(\rho)}{\partial \rho} \hskip 0.3 cm , \hskip 0.5cm
\Gamma_B(p=0;\rho) = \frac{\partial^2 V_\kappa(\rho)}{\partial \rho^2} .
\end{equation}

Now, the effective potential satisfies an exact flow equation which can
be deduced  from that for the effective action, eq.~(\ref{NPRGeq}),
when restricted to constant fields. It reads
\begin{equation}\label{eqforV}
\partial_t V_\kappa(\rho)=\frac{1}{2}\int
\frac{d^dq}{(2\pi)^d}
\partial_t R_\kappa(q) \Big\{ (N-1) G_T(q;\rho) + G_L(q;\rho)
\Big\} = \frac{1}{2} \Big\{ (N-1) I_{d;T}^{(1)}(\rho;\kappa) + I_{d;L}^{(1)}(\rho;\kappa) \Big\} .
\end{equation}
(In the last term of the equation, we made a slight abuse of language with respect to the definition of eq. (\ref{defI}): the function $I_{d;\alpha}^{(1)}$ has a unique index $\alpha$ because it contains a unique propagator.)

According to eq. (\ref{zeromomenta}), the second derivative  of eq. (\ref{eqforV}) with respect to the
background field gives a flow equation for $\Gamma^{(2)}_{ab}
(p=0;\rho)$. Now, this  equation  does not coincide with
eq.~(\ref{2pointclosed}) with $p=0$: indeed, in contrast
to eq.~(\ref{2pointclosed}), the vertices in the equation deduced
from eq.~(\ref{eqforV}) keep all their $q$-dependence ($q$ being the momentum in the loop integral); in other words, it is a more precise equation. There is therefore an apparent
inconsistency between eqs. (\ref{2pointclosed}), (\ref{zeromomenta}) and (\ref{eqforV}). However,  this can be easily
solved. To do so, it is convenient to treat separately the zero momentum ($p=0$) and the
non-zero momentum ($p \ne 0$) sectors (for a further discussion, see \cite{BMWnum}).

Let us then define:
\begin{align}
\Delta_A(p; \rho;\kappa) \equiv & \; \Gamma_A(p; \rho;\kappa) - p^2  -
\Gamma_A(p = 0; \rho;\kappa) \label{defdelta_a_2} \\
\Delta_B(p; \rho;\kappa) \equiv & \; \Gamma_B(p;
\rho;\kappa) - \Gamma_B(p = 0; \rho;\kappa) \label{defdelta_b_2} .
\end{align}
The flow equations for $\Delta_A$ and $\Delta_B$ easily follow from those for $\Gamma_A$ and $\Gamma_B$:
\begin{equation}\label{flowdelta_a}
\partial_t \Delta_A(p; \rho) =  \partial_t
\Gamma_A(p; \rho;\kappa) - \partial_t\Gamma_A(p = 0; \rho;\kappa)
\end{equation}
\noindent and equivalently for $\Delta_B(p;\rho;\kappa)$.

The procedure we shall consider in this paper consists then in solving simultaneously the 3 flow equations for $V_\kappa(\rho)$ (eq. (\ref{eqforV})), $\Delta_A(p;\rho;\kappa)$ and $\Delta_B(p;\rho;\kappa)$, and then get
$2$-point functions through
\begin{align}
\Gamma_A(p; \rho;\kappa) & = p^2 + \Delta_A(p; \rho;\kappa) + \frac{\del
V_\kappa(\rho)}{\del \rho} \label{defdelta_a} \\
\Gamma_B(p; \rho;\kappa) & =
\Delta_B(p; \rho;\kappa) + \frac{\del^2 V_\kappa(\rho)}{\del \rho^2} .
\label{defdelta_b}
\end{align}
The initial conditions for the flow are $V_\Lambda(\rho) = r_{\Lambda} \rho + (u/6)\rho^2$ (see eq.
(\ref{Sclassical})) for the potential, while those for both $\Delta$ functions are equal to $0$, as the classical momentum dependence is explicitly taken out in their definition (see eqs. (\ref{defdelta_a_2}) and (\ref{defdelta_b_2})).
Notice that, proceeding in this way, not only we maintain the validity of the relationship given by eq. (\ref{zeromomenta}), but we also gain more accuracy in the description of the 2-point function: indeed, its momentum independent part is now described with a higher precision, while the approximation introduced in \cite{BMW}, which is used in this paper, only affects its $p$ dependence.

\section{Resolution Strategies}

\label{strategies}

Although the flow equations for $V_\kappa(\rho)$, $\Delta_A(p;\rho;\kappa)$ and
$\Delta_B(p;\rho;\kappa)$ can in principle be solved numerically, this nonetheless constitutes a rather cumbersome task. The reason is twofold. First, what we called $I$ and
$J$ are in fact functionals of the solution $\Gamma^{(2)}(p,\rho;\kappa)$; this
complicates the possible integration strategies. Second,
notice that different values of $p$ are
coupled through the propagators $G_{\alpha}(p+q;\kappa)$ which enter in the
calculations of the $J$ functions; this demands in principle
the simultaneous solution of the equations for all $p$.

Nevertheless, we shall show that within a simple, and yet accurate,
further approximation, our flow equations become
numerically simpler. In this section we shall in fact discuss two possible approximation strategies.

\subsection{First Level of Approximation: Strategy I}

\label{strategyone}

In order to simplify the above mentioned issues, and to bring down the numerical
effort necessary to solve our set of NPRG equations, it is possible to perform a further approximation (see \cite{BMWnum}, where the same approximation was used in the $N=1$ case and an assessment of its accuracy was done).

Consider first the function $I_{d,\alpha \beta}^{(n)}(\rho;\kappa)$, which
does not depend on $p$. The smoothness of the $n$-point functions
and the fact that integrals are dominated by the domain $q \lesssim
\kappa$, suggest to perform in the propagators of the
right-hand-side of eq.~(\ref{defI}) an approximation similar to that
applied to the other $n$-point functions, i.e., to set $q=0$. However,
as already said in the previous section, in order to maintain both the exact one-loop and large $N$ properties of the flow
equations,  one cannot simply set $q=0$ in the whole propagator: rather,
one needs a momentum dependence recovering that of the free
propagators in the $\kappa \to \Lambda$ limit. Thus, we shall use for  the propagators entering the
calculation of $I^{(n)}_{d;\alpha\beta}(\rho;\kappa)$ the following
approximate forms
\begin{align} G_T^{-1}(q;\rho;\kappa)
& \approx Z_\kappa q^2 + \Gamma_A(q=0;\rho;\kappa) + R_\kappa(q),
\label{propGT} \\
G_L^{-1}(q;\rho;\kappa) & \approx Z_\kappa q^2 +
\Gamma_A(q=0;\rho;\kappa) + 2 \rho \Gamma_B(q=0;\rho;\kappa)+ R_\kappa(q),
\label{propGL}
\end{align}
\noindent where
\begin{align}
\label{def-Zk}
Z_\kappa \equiv\left.
{\frac{\partial \Gamma_\kappa^{(2)}}{\partial
q^2}}\right|_{q=0,\rho=\rho_0}.
\end{align}
As we shall see in the following, the presence of the $Z_\kappa$ factor is needed in order to preserve scaling properties.
As it is well known \cite{Morris94c} $\left.\partial
\Gamma^{(2)}(q;\rho)/\partial q^2\right|_{q=0}$ depends weakly on
$\rho$. Accordingly, one expects $Z_\kappa$ to depend weakly on the
value chosen for $\rho_0$. As argued in \cite{BMWnum}, the choice
$\rho_0=0$ is here the simplest one. With the propagators of eqs. (\ref{propGT})
and (\ref{propGL}), and the choice of the regulating function \cite{Litim}
\begin{equation}\label{reg-litim}
R_\kappa(q) = Z_\kappa
(\kappa^2 - q^2) \Theta(\kappa^2 - q^2) ,
\end{equation}
the function $I_{d;\alpha\beta }^{(n)}(\rho;\kappa)$ can be calculated analytically:
\begin{equation} \label{In-anal}
I_{d,\alpha
\beta}^{(n)}(\rho;\kappa) = 2 K_d
\frac{\kappa^{d+2-2n}}{Z_\kappa^{n-1}}
\left(1-\frac{\eta_\kappa}{d+2}\right) \frac{1}{(1+\hat
m^2_{\alpha}(\rho))^{n-1} (1+\hat m^2_{\beta}(\rho))}.
\end{equation}
In this expression,
\begin{equation}\label{defetak}
\eta_\kappa\equiv-\kappa\partial_\kappa \ln Z_\kappa
\end{equation}
is the running anomalous dimension and
\begin{align}
\hat m^2_T(\rho;\kappa) & \equiv
\frac{\Gamma_A(q=0;\rho;\kappa)}{\kappa^2
Z_\kappa} = \frac{V_\kappa'(\rho)}{\kappa^2 Z_\kappa} \label{defm2T} \\
\hat m^2_L(\rho;\kappa) & \equiv \frac{\Gamma_A(q=0;\rho;\kappa) + 2\rho
\Gamma_B(q=0; \rho;\kappa)}{\kappa^2 Z_\kappa} = \frac{V_\kappa'(\rho) + 2\rho
V_\kappa''(\rho)}{\kappa^2 Z_\kappa}, \label{defm2L}
\end{align}
are dimensionless, field-dependent, effective masses.
Above, $K_d$ is a number resulting from angular integration,
$K_d^{-1}\equiv d\; 2^{d-1}\; \pi^{d/2} \; \Gamma(d/2)$ (e.g.,
$K_3=1/(6\pi^2)$). Notice that, for $d>2$, $I_{d, \alpha \beta}^{(2)}(\kappa;\rho) \to 0$ when $\kappa \to 0$.

As for the function $J_{d, \alpha \beta}^{(n)}(p;\rho;\kappa)$, we
shall calculate it in a similar way. To do so, let us notice that
the propagator
$G_{\alpha}(p+q;\rho)$ in eq.~(\ref{defJ}) is small as soon as
$p/\kappa$ is large; one can verify  that the function
$J^{(3)}_{d,\alpha\beta}(p;\rho;\kappa)$ vanishes approximately as
$\kappa^2/p^2$ for large values of $p/\kappa$. Thus, in the region where $J^{(3)}_{d, \alpha
\beta}(p;\rho;\kappa)$ has a non-negligible value, one can
assume $p\simle \kappa$ and then
use for
$G_\alpha(p+q;\rho)$ an expression  similar to that of eqs. (\ref{propGT}) or
(\ref{propGL}), namely
\begin{align}
G_T^{-1}(p+q;\rho;\kappa) & \approx Z_\kappa (p+q)^2 +
\Gamma_A(q=0;\rho;\kappa) + R_\kappa(p+q), \label{propGTp+q} \\
G_L^{-1}(p+q;\rho;\kappa) & \approx Z_\kappa (p+q)^2 + \Gamma_A(0;\rho;\kappa)
+ 2\rho \Gamma_B(q=0;\rho;\kappa)+ R_\kappa(p+q), \label{propGLp+q}
\end{align}

One can then calculate the functions $J^{(3)}_{d,\alpha\beta}(p;\rho;\kappa)$ analytically (in $d=3$ and with the regulator of eq. (\ref{reg-litim})). The resulting expressions
are more complicated than those for $I_{d,\alpha \beta}^{(2)}(\rho;\kappa)$,
eq.~(\ref{In-anal}). They are given in appendix \ref{functionf}.
 Observe that the regulator in eq.~(\ref{reg-litim}) is
not analytic at $q = \kappa$. This generates non analyticities in
$J^{(3)}_{d,\alpha \beta}(p;\rho;\kappa)$; but
 these occur  only in the third derivative with respect
  to $p$, at $p=0$ and at $p=2\kappa$
  (cf., e.g.,  the odd powers of $\bar p$ in eqs.~(\ref{Jlarge}-\ref{largeJ4})),
   and they  play no role at the present level of approximation.

Finally, for the last term in eq. (\ref{gamma_b}), it is possible to use the same approximation procedure described above; doing so,  $\Gamma_B(q;\rho) \simeq \Gamma_B(q=0;\rho) = V''(\rho)$, and the term is then just proportional to the sum of two $I^{(3)}$ functions.

With the approximation just discussed, both $I_{d,\alpha\beta}^{(n)}(\rho;\kappa)$ and $J^{(3)}_{d,\alpha \beta}(p;\rho;\kappa)$ become explicit functions of the
potential $V_\kappa(\rho)$ and the field renormalization constant $Z_\kappa$ (or, equivalently, $\eta_\kappa$, see eq, (\ref{defetak})). From now on, we shall denote the range of momenta $p \simle \kappa$, which is described by quantities as
$V_\kappa(\rho)$ and $Z_\kappa$ (or $\eta_\kappa$), as the ``$p=0$ sector'' of the theory.

To finish the description of the calculation procedure it is then
necessary to make explicit how to solve these two flow equations.
The simplest way is the usual procedure of DE. Specific details are
presented in Appendix \ref{LPAp}. Here we shall just quote three
ingredients which are relevant for our present discussion. First
notice that, within DE, the integral $I_{d;\alpha}^{(1)}$ appearing
in the potential flow equation is calculated using the approximate
propagators from eqs. (\ref{propGT}) and (\ref{propGL}); i.e., the
integral is given by eq. (\ref{In-anal}). We shall be  back to this
point in the next sub-section. Second, in order to get the proper
scaling behavior of $\Gamma^{(2)}(p;\rho;\kappa)$ at small momenta,
we need the flow equation for $Z_\kappa$ to be consistent with the
approximate eq.~(\ref{2pointclosed}) for the 2-point function. This
is achieved by extracting $Z_\kappa$ from the flow equation of
$\lim_{p\to 0} \Delta_A(p;\rho=0;\kappa)/p^2$, which follows from
eq.~(\ref{2pointclosed}), and invoking eq.~(\ref{def-Zk}). Details
can be found in appendix \ref{LPAp}. Third, let us notice that, in
fact, the flow equation for the potential is qualitatively different
than  those for $\Delta_A(p;\rho;\kappa)$ and
$\Delta_B(p;\rho;\kappa)$. Indeed, these depend on a dimensionful
quantity $p$; thus, when $\kappa \to 0$, the corresponding flow stops,
giving a finite value for both 2-point functions, which depends on
$p$. On the other side, at criticality, the potential only depends
on the dimensionful physical variable $\rho$, whose relevant values
shrink to zero when $\kappa \ll u$; the system is then characterized
by scale invariance. In order to correctly parametrize this
property, it is convenient to work with dimensionless variables
\begin{equation}\label{adim}
\tilde{\rho}  \equiv K_d^{-1} Z_\kappa \kappa^{2-d} \rho \hskip 0.3cm , \hskip 0.5cm
 v_\kappa(\tilde \rho)  \equiv K_d^{-1} \frac{V_\kappa(\rho)}{\kappa^d}
\end{equation}
Doing so, the dimensionless potential
$v_\kappa(\tilde \rho)$ approaches a non trivial fixed point form when $\kappa \ll u$.

The strategy to solve the eq.~(\ref{2pointclosed}) for the flow of
the $2$-point function consists then in two steps: one first solves
the $p=0$ sector to get $v_\kappa(\tilde{\rho})$ and $\eta_\kappa$;
in doing so, the bare mass is adjusted in order to reach the IR fixed point. Second, for each value of $p$, one solves the flow equations for the $\Delta$'s, where the kernels
$I_d^{(n)} (\rho;\kappa)$ and $J_d^{(3)}(p;\rho;\kappa)$
are explicit functions of $v_\kappa(\rho)$, $Z_\kappa$ and $\eta_\kappa$.
The problem of finding the $2$-point function of the $O(N)$
symmetric scalar field is thus reduced to the solution of a system
of partial differential equations with parameter $p$, which can be solved separately
for each value of $p$, and which does not involve a numerical
effort greater than that required in usual DE calculations.

\subsection{Improved Approximation: Strategy II}

\label{improved}

In section \ref{introduction} we recalled an interesting  property of the approximation scheme introduced in \cite{BMW}: when solving the flow equation of the $2$-point function at the LO of the scheme, the effective potential one gets is exact at $2$ loops. Nevertheless, when solving eq. (\ref{eqforV}) using the propagators described in the previous subsection, this $2$-loop exactness is lost. We shall present now a simple improvement in the procedure proposed in subsection \ref{strategyone} in order to recover the $2$-loop expression for the potential. This should bring a better description of the $p=0$ sector of the model, which would be particularly useful for the determination of critical exponents.

In order to do so, let us briefly remind the origin of the $2$-loop exactness. It
exploits the fact that only one-loop diagrams contribute to the flow
(see for example eqs.~(\ref{NPRGeq}), (\ref{gamma2}), (\ref{eqforV}), or
figure \ref{diagrams}).
Accordingly, the $2$-point function gets exact at one loop, provided its flow is calculated with quantities
which are exact in the classical limit. This is respected not only within the approximate eq. (\ref{2pointclosed}), but also after the extra approximation in the propagators introduced in subsection \ref{strategyone}. As for the potential, eq. (\ref{eqforV}) gives indeed an exact expression at $2$-loops only if the flow is calculated with  quantities exact at $1$-loop. This is in principle the case for the equations here considered. Nevertheless, the extra approximation introduced in subsection \ref{strategyone}, when applied to the potential flow equation, violates this property. Fortunately, this can be swiftly solved, and with a low numerical cost.

This improvement can be achieved by numerically integrating the function $I^{(1)}$ appearing in the flow equation for the potential, without using the approximate propagator of eqs. (\ref{propGT}) and (\ref{propGL}) but, instead, the numerical solutions for $\Delta_A$ and $\Delta_B$:
\begin{equation}\label{eqforVimpr}
\begin{split}
\partial_t V_\kappa(\rho) & =\frac{1}{2}\int
\frac{d^dq}{(2\pi)^d} \partial_t R_\kappa(q) \bigg(
(N-1) \frac{1}{q^2 + \Delta_A(q;\rho) + V'(\rho) + R_\kappa(q)} \\
& \quad + \frac{1}{q^2 + \Delta_A(q;\rho) + 2\rho \Delta_B(q;\rho) +
V' + 2\rho V'' + R_\kappa(q)} \bigg).
\end{split}
\end{equation}
One thus needs to simultaneously solve the flow equations for $V_\kappa$, $\Delta_A$ and $\Delta_B$.

Notice that the approximate propagators, and thus the analytic expressions for $J$, from appendix \ref{functionf}, and for $I^{(2)}$ and $I^{(3)}$, from eq. (\ref{In-anal}), can still be used in the flow equations for
$\Delta_A$ and $\Delta_B$, without loosing their $1$-loop exactness.

In eq. (\ref{eqforVimpr}), angular integration can be done
analytically reducing the problem to a numerical integration over
one single variable, $|q|$. Accordingly, the procedure does not
introduce too much extra complexity in the algorithm. However, an
important subtlety arises: due to the regulator, the integrand of
$I^{(1)}$ takes non-negligible values only in the range $|q| \lesssim
\kappa$. Thus, with a fixed grid in $q$, as $\kappa$ goes to the
physical value $\kappa=0$, the number of points in $q$ for
performing a numerical integration in eq. (\ref{eqforVimpr}) would
dwindle very rapidly.

This apparent difficulty is cured by working with fixed values of $q/\kappa$, i.e., solving the $\Delta$ functions flow equations within a grid for fixed values of $\tilde q \equiv q/\kappa$ with $\tilde q < \tilde q_{max}$; values of $\tilde q_{max} \sim 3-4$ turn out to be large enough. Nevertheless, due to their definition (see eqs. (\ref{defdelta_a_2}) and (\ref{defdelta_b_2})), when $\kappa \to 0$, all functions in the grid would vanish.
As usually done for the potential (see eq.~(\ref{adim})), this difficulty is simply solved working with dimensionless variables:
\begin{align}\label{adim2}
 \tilde{\Delta}_A(\tilde{p}; \tilde{\rho}) & \equiv  \frac{\Delta_A(p;\rho)+p^2}{\kappa^2 Z_\kappa}, & \tilde{\Delta}_B(\tilde{p}; \tilde{\rho}) & \equiv
  \frac{\Delta_B(p;\rho)}{\kappa^{4-d} Z_\kappa^2 K_d   ^{-1}} .
\end{align}
At criticality, these quantities reach finite values in the $\kappa \to 0$ limit, this limit depending on the value of $\tilde p = p/\kappa$. These functions $\tilde \Delta_A$ and $\tilde\Delta_B$ are precisely the quantities entering the integrand of the flow equation for the dimensionless potential $v_\kappa$ (which follows from eqs. (\ref{adim}) and (\ref{eqforVimpr})).

As a final remark concerning the $p=0$ sector, notice that we now have many fixed point flow equations: those for the dimensionless potential $v_\kappa$ and for $\eta_\kappa$, and those for the dimensionless $\tilde \Delta$ functions (in fact, two equations for each value of $\tilde q$ on the grid). As we are dealing with a Wilson-Fisher fixed point, the flow has only one unstable direction; thus once only one bare parameter is fined tuned (here, the bare mass), the complete set of equations should reach the fixed point. However, handling the flow equations for $\tilde \Delta_A$ and $\tilde \Delta_B$ turns out to be a hard numerical task. For reasons of numerical stability, it proves useful to introduce
the auxiliary variables
\begin{align}\label{yes}
Y_A(\tilde{p};\tilde{\rho}) & \equiv
\frac{\tilde{\Delta}_A(\tilde{p};\tilde{\rho})}{\tilde{p}^2} , &
Y_B(\tilde{p};\tilde{\rho}) & \equiv
\frac{\tilde{\Delta}_B(\tilde{p};\tilde{\rho})}{\tilde{p}^2} .
\end{align}

Flow equations for $Y_A$ and $Y_B$ are trivially derived. The departing point is the flow equations for
$\Delta_A$ and $\Delta_B$ we used in the previous subsection, i.e., those with the approximated analytic expressions for the functions $J^{(3)}$, $I^{(2)}$ and $I^{(3)}$. Then, making use of
eqs. (\ref{adim2}) and (\ref{yes}), a straightforward calculation yields:
\begin{align}\label{flowYa}
  \partial_t Y_A(\tilde{p}; \tilde{\rho}) &= \eta Y_A(\tilde{p}; \tilde{\rho}) + \tilde{p} \frac{\partial
  Y_A}{\partial \tilde {p}}(\tilde{p}; \tilde{\rho}) + (d - 2 + \eta) \tilde{\rho} Y_A' + 2 \left(1
-\frac{\eta}{d+2} \right) \nonumber \\
& \times \Bigg[ - \frac{1}{\tilde{p}^2}\frac{2\tilde{\rho}}{(1+w+2\tilde{\rho}w')^2} \frac{w'^
2}{1+w}
-  \frac{1}{\tilde{p}^2}\frac{2\tilde{\rho}}{(1+w)^2} \frac{w'^2}{1+w+2\tilde{\rho}w'} \nonumber \\
& - \frac{1}{2} \frac{(Y_A'(\tilde{p}; \tilde{\rho}) +
2\tilde{\rho}Y_A''(\tilde{p}; \tilde{\rho}))}{(1 +w
    +2\tilde{\rho}w')^2}
- \frac{1}{2} \frac{((N-1) Y_A'(\tilde{p}; \tilde{\rho}) + 2 Y_B(\tilde{p}; \tilde{\rho}))}{(1+w)^2} \Bigg] \nonumber \\
& + 2\tilde{\rho} \tilde{J}_{LT}(\tilde{p}; \tilde{\rho})
\big(Y_A'^2(\tilde{p}; \tilde{\rho}) \tilde{p}^2 + 2
 Y_A'(\tilde{p}; \tilde{\rho}) w' + \frac{w'^2}{\tilde{p}^2}\big) \nonumber \\
& + 2\tilde{\rho} \tilde{J}_{TL}(\tilde{p}; \tilde{\rho})  \big(
Y_B^2(\tilde{p}; \tilde{\rho}) \tilde{p}^2 + 2
Y_B(\tilde{p}; \tilde{\rho})w' + \frac{w'^2}{\tilde{p}^2}\big)
\end{align}
\begin{align}\label{flowYb}
\partial_t Y_B(\tilde{p};\tilde{\rho}) &= (d
- 2 + 2 \eta) Y_B(\tilde{p}; \tilde{\rho}) + \tilde{p}
\frac{\partial Y_B}{\partial \tilde{p}}(\tilde{p};
\tilde{\rho}) + (d-2+\eta) \tilde{\rho} Y_B'(\tilde{p};
\tilde{\rho}) \nonumber \\
& + 2\left(1 - \frac{\eta}{d+2} \right) \Bigg[ \frac{(N-1)
w'^2}{\tilde{p}^2(1+w)^3} + \frac{( 9 w^2 + 12 \tilde{\rho} w' w'' + 4
\tilde{\rho}^2 w''^2)}{\tilde{p}^2 (1+w+2\tilde{\rho}w')^3} \nonumber \\ &
- \frac{1}{\tilde{p}^2}\frac{1}{(1+w+2\tilde{\rho}w')^2} \frac{w'^2}{1+w}  -
\frac{1}{\tilde{p}^2}\frac{1}{(1+w+2\tilde{\rho}w')} \frac{w'^2}{(1+w)^2} \nonumber
\\& - \frac{1}{2}
\frac{(N-1)}{(1+w)^2} Y_B'(\tilde{p}; \tilde{\rho}) -
  \frac{1}{2} \frac{(5 Y_B'(\tilde{p};
\tilde{\rho}) + 2\tilde{\rho} Y_B''(\tilde{p};
\tilde{\rho}))}{(1+w+2\tilde{\rho}w')^2}
    \nonumber \\
& + \left( \frac{1}{1+w+2\tilde{\rho}w'} + \frac{1}{1+w} \right)
\frac{1}{1+w}
    \frac{Y_B(\tilde{p};
\tilde{\rho}) w'}{1+w+2\tilde{\rho}w'} \Bigg] \nonumber \\
& + (N-1) \tilde{J}_T(\tilde{p}; \tilde{\rho}) \big(
  Y_B^2(\tilde{p};
\tilde{\rho}) \tilde{p}^2 + Y_B(\tilde{p}; \tilde{\rho}) w' +  \frac{w'^2}{\tilde{p}^2}\big) \nonumber \\ & + \tilde{J}_L(\tilde{p}; \tilde{\rho}) \Big\{Y_A'^2(\tilde{p};
\tilde{\rho}) \tilde{p}^2 + 4 Y_B^2(\tilde{p}; \tilde{\rho}) \tilde{p}^2 + 6
Y_A'(\tilde{p}; \tilde{\rho}) w'  + \frac{9
  w'^2}{\tilde{p}^2} \nonumber \\ & + 4 Y_A'(\tilde{p};
\tilde{\rho})  Y_B(\tilde{p} \tilde{p}^2;
\tilde{\rho}) + 12 Y_B(\tilde{p}; \tilde{\rho}) w' +
4\tilde{\rho} (Y_A'(\tilde{p}; \tilde{\rho})
Y_B'(\tilde{p}; \tilde{\rho}) \tilde{p}^2 \nonumber \\ &
+ Y_A'(\tilde{p}; \tilde{\rho}) w'' +3
Y_B'(\tilde{p}; \tilde{\rho}) w' + \frac{3 w'
  w''}{\tilde{p}^2} + 2 Y_B(\tilde{p};
\tilde{\rho}) Y_B'(\tilde{p};
\tilde{\rho}) \tilde{p}^2 \nonumber \\
& + 2 Y_B(\tilde{p}; \tilde{\rho}) w'' + 2
Y_B'(\tilde{p}; \tilde{\rho}) w') + 4\tilde{\rho}^2
\big( Y_B'^2(\tilde{p}; \tilde{\rho})) \nonumber \\
& + 2
  Y_B'(\tilde{p};
\tilde{\rho}) w'' + \frac{w''^2}{\tilde{p}^2}\big)\Big\} - \tilde{J}_{LT}(\tilde{p};
\tilde{\rho}) \big(
  Y_A'^2(\tilde{p};
\tilde{\rho}) \tilde{p}^2 + 2  Y_A'(\tilde{p};
\tilde{\rho}) w' \nonumber \\
& + \frac{w'^2}{\tilde{p}^2}\big) -  \tilde{J}_{TL}(\tilde{p}; \tilde{\rho})
\big(Y_B^2(\tilde{p}; \tilde{\rho}) \tilde{p}^2 + 2
  Y_B(\tilde{p};
\tilde{\rho}) w' + \frac{w'^2}{\tilde{p}^2}\big)
\end{align}

Notice that the process of going to dimensionless variables brought
into play derivatives with respect to $\tilde{p}$. Above, we
introduced the dimensionless expression $\tilde{J}_{\alpha
\beta}(\tilde{p}; \tilde{\rho})$ of the function $J$, which is given
in appendix \ref{functionf}. The ($\kappa=\Lambda$) initial values
for these functions are $Y_A=1$, $Y_B=0$.

In summary, according to this second strategy to get the $2$-point
functions of the $O(N)$ model, one also proceeds in two steps.
First, one fixes the $p=0$ sector, which now demands the
simultaneous solution of the full flow equation for the potential
(eq. (\ref{eqforVimpr})) together with those for $Y_A$ and $Y_B$
(eqs. (\ref{flowYa}) and (\ref{flowYb})) for a limited number of
values of $\tilde q < \tilde q_{max}$. Proceeding this way one
fine-tunes the bare mass in order to get the infrared fixed point.
As a second step, one solves the flow equations for dimensionful
$\Delta_A$ and $\Delta_B$ (the same one as those of the previous
subsection) for any desired value of the external momenta $p$.
Within this second strategy, once the $p=0$ sector is solved, the
$p\neq 0$ sector can still be treated separately for each value of
$p$.

Using the alternative idea just presented, one hopes to have a better description of the $p=0$ sector, and thus, to get an improvement in critical exponents and infrared properties of the model. Moreover, one can also hope to
retrieve better results at least in the cross-over region between IR and UV
behavior.

\section{Numerical Results and Discussion}

\label{results}

We now turn to the numerical solutions obtained within both of the methods we have
discussed in the previous section. We shall consider here the
properties of the system near or at criticality, and we shall study
the $d = 3$ case. Our goal here is twofold: first, to asses the
quality of the approximation scheme at all ranges of momenta and to
verify what is the effect of the improvement discussed in the last
subsection. Second, since we are working here only with the LO of
the approximation method proposed in \cite{BMW} (and, moreover, using approximate propagators) we aim to compare our
results with those already existing in the literature for the $O(N)$ model.

It is possible to distinguish three momentum regions: the IR sector
($p \ll u$), the UV sector ($p\gg u$) and the intermediate
cross-over sector. The first region is dominated by scaling
properties and has been studied using many different methods with
high precision. In the next subsection we shall present our
predictions for this regime, paying particular attention to scaling
properties and comparing our results with those existent in the
literature. As for the UV regime, it can also be described with high
precision within perturbative calculations. Finally, the crossover
region is known with much less accuracy; it is therefore the main
yield of this work. Both the UV and the crossover regimes shall be
discussed in the second part of this section.

\subsection{Scaling properties}

When $\kappa \ll p \ll u$, we expect $\Delta_A(p;\kappa)$ to behave as
\beq\label{delta-scaling}
p^2+\Delta_A(p;\kappa) = Ap^{2-\eta^*}
\eeq
where $\eta^*$ is the anomalous
dimension, i.e., the fixed point value of $\eta_\kappa$. As the explicit $p$-dependence of the $2$-point function is very hard to obtain, in the literature, the usual way to determine $\eta^*$ is to extract it from the
$\kappa$ dependence of $Z_\kappa$ (see eq.~(\ref{defetak})):
\beq
Z_\kappa \propto \kappa^{- \eta^*},
\eeq
when $\kappa \ll u$. Nevertheless, although these two extractions should in principle lead to the same result \cite{Blaizot05}, when approximating the flow equations this property can be lost. This is for example the case in DE which, at criticality, describes physical quantities only at $p = 0$; not even the $\kappa \ll p \ll u$ regime is correctly described. We have checked that, for all the solutions we get, we always have the same result within the two extraction methods.

\begin{figure}[t]
\begin{center}
\includegraphics*[scale=0.28,angle=-90]{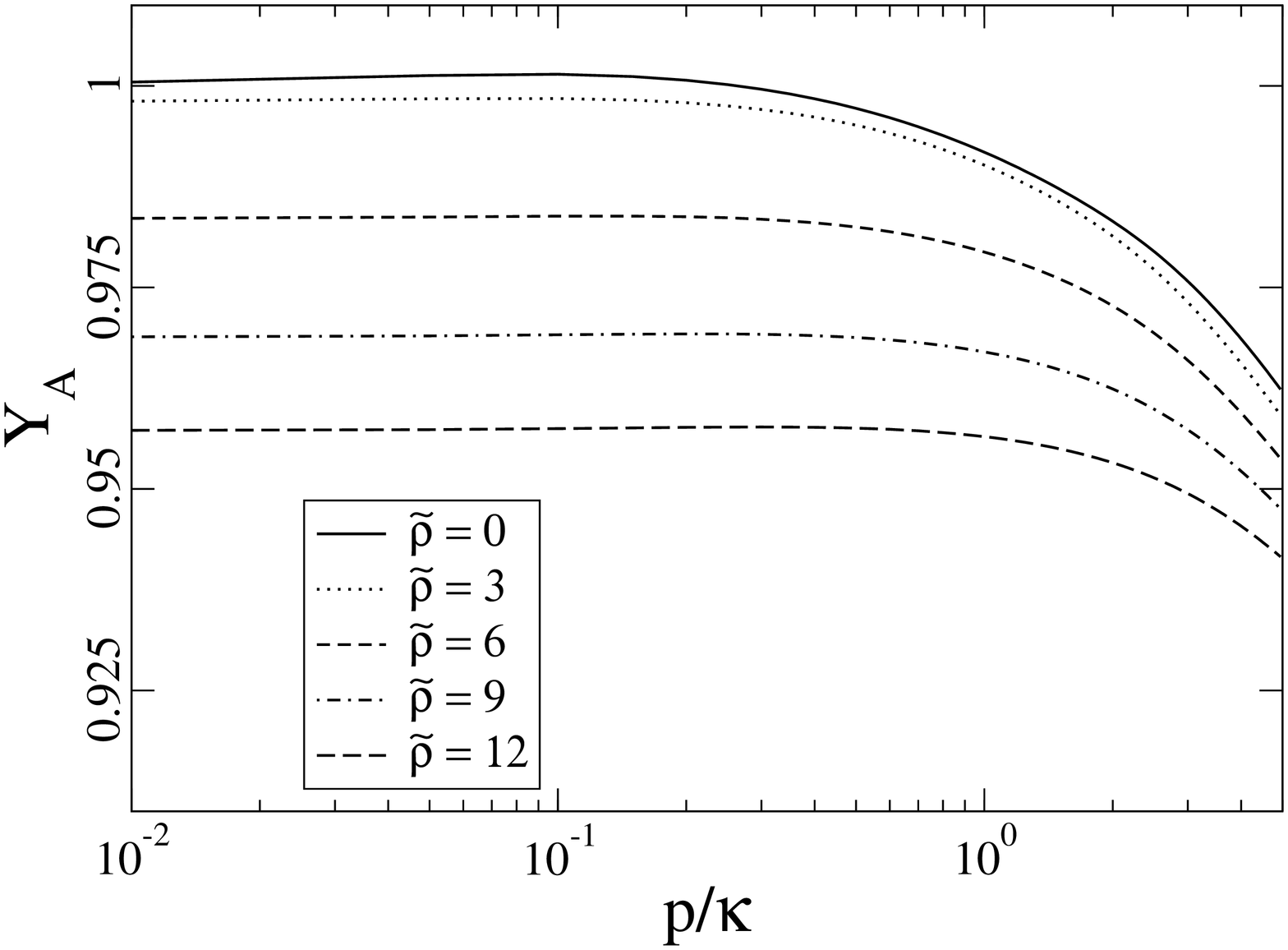} \includegraphics*[scale=0.28,angle=-90]{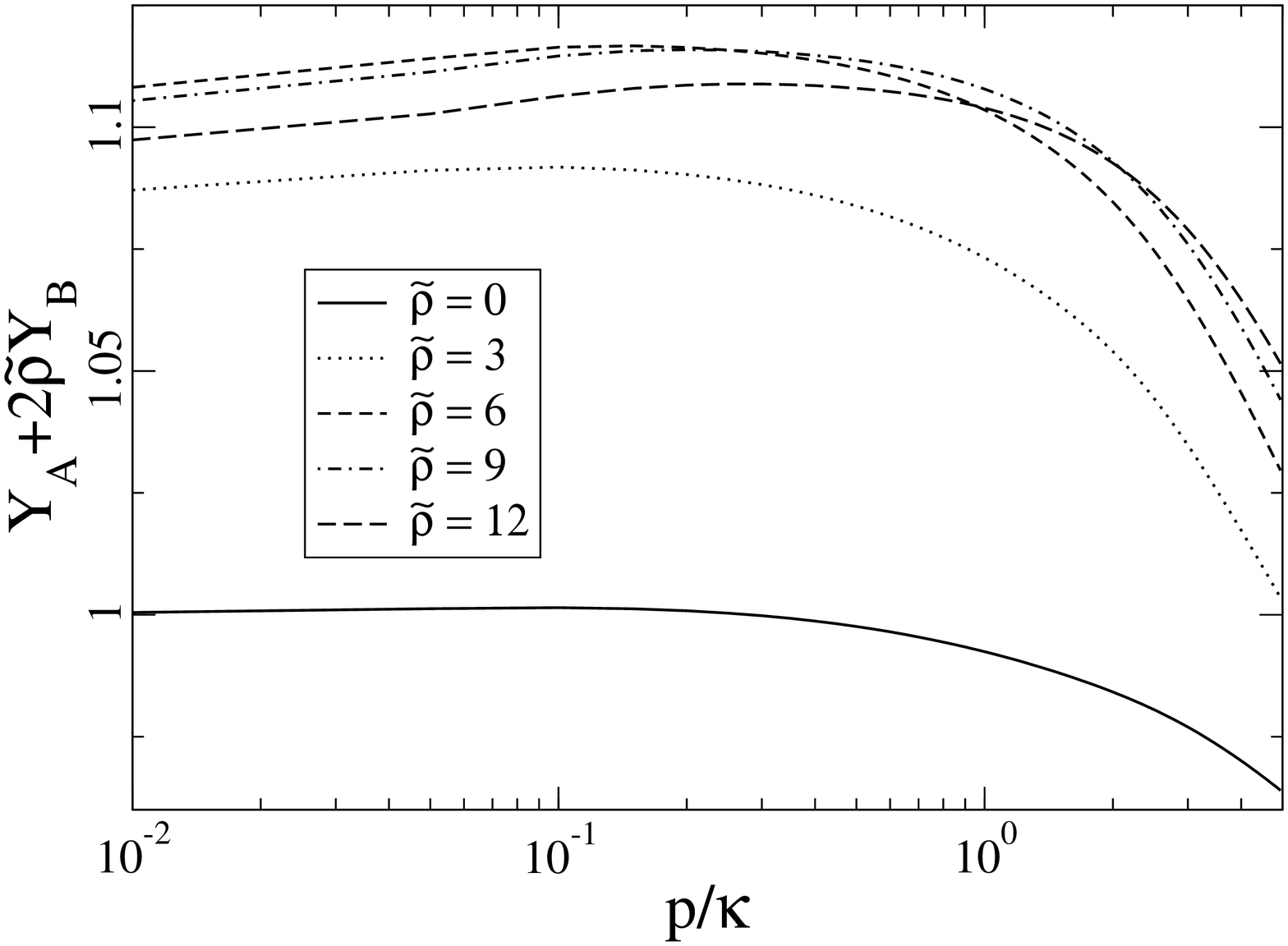}
\end{center}
\caption{\label{figYA} The fixed point form for the functions $Y_A(\tilde p, \tilde \rho)$ and $Y_A(\tilde p, \tilde \rho) + 2 \tilde \rho Y_B(\tilde p, \tilde \rho)$, in $d=3$, $N=2$, at
criticality, as a function of $\tilde{p}$, for various values of $\tilde \rho$.}
\end{figure}

A more stringent test of scaling is given by the study of the
dimensionless quantities \beq \frac{\Delta_A(p;\tilde
\rho;\kappa)+p^2}{p^2 Z_\kappa}, \hskip 3 cm \frac{\Delta_A(p;\tilde
\rho;\kappa)+2\rho\Delta_B(p;\tilde \rho;\kappa)+p^2}{p^2 Z_\kappa}
\eeq which, in the scaling regime ($p, \kappa\ll u$) should be
functions of only $p/\kappa$ and $\tilde\rho$. In fact, according to
eqs. (\ref{adim2}) and (\ref{yes}) these should be the scaling
functions $Y_A(\tilde{p};\tilde\rho)$ and
$Y_A(\tilde{p};\tilde\rho)+ 2\tilde\rho Y_B(\tilde p;\tilde \rho)$.
We have numerically checked that our results verify this property.
In figure \ref{figYA} we show these scaling functions, as a function
of $\tilde p$, for various values of $\tilde\rho$ and $N=2$. Similar results are found for other values of $N$.

Notice that, by definition of $Z_\kappa$ (see eq.~(\ref{def-Zk})),
the function $Y_A(\tilde p=0, \tilde \rho=0)=1$.  A non-trivial fact,
shown by figure \ref{figYA}, is that
both the functions $Y_A(\tilde p, \tilde \rho)$ and $Y_A(\tilde p,
\tilde \rho)+2 \tilde \rho Y_B(\tilde p, \tilde \rho)$
are well approximated by unity for all $\tilde{p} \lesssim
1$ and all relevant values of $\tilde \rho$ (as well known
\cite{Berges00}, the latter are those of the order of the
minimum of the potential which, within our normalization, runs from
$\tilde \rho\sim N+2$, when $\kappa = \Lambda$,
to $\tilde \rho \sim N$, when $\kappa \to 0$). This is exactly what we
assumed within the two numerical
approximations introduced in the last section. In other words, the
figure shows that, even in the deep IR, the approximated
propagators of eqs. (\ref{propGT}) and (\ref{propGL}) are indeed
accurate. In ref.~\cite{BMWnum}, where the
same approximation is done in order to solve the $N=1$ case, its effect
on the functions $I$ and $J$ is shown to
be small. There, only the longitudinal propagator, i.e., that given by
eq.~(\ref{propGL}), plays a role (remember
that the term including the transverse propagator is proportional to
$N-1$). Figure \ref{figYA} shows that the
approximation is even better in the transverse case; accordingly, when
going to large values of $N$, one expects
the approximations introduced in the last section to become better. As
we shall see, this is confirmed by our
numerical results.

\begin{figure}[t]
\begin{center}
\includegraphics*[scale=0.4,angle=-90]{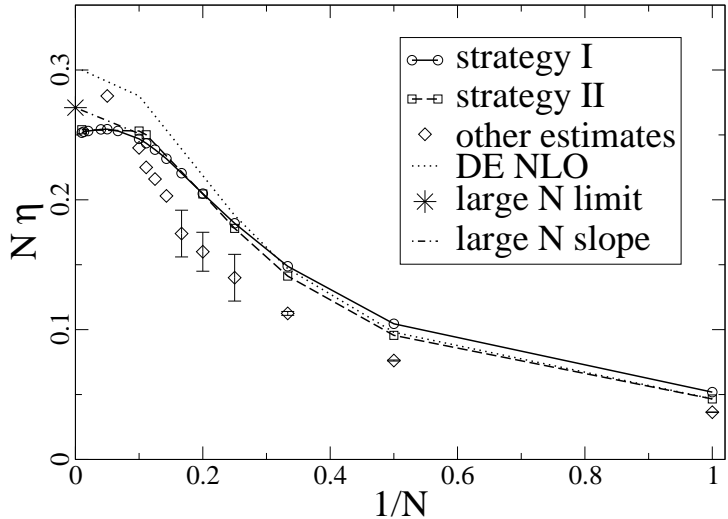}
\end{center}
\caption{ \label{graf_etas} $N\eta$ as a function of $1/N$ for various model calculations, together with the $1/N$ analytical results. }
\end{figure}

Our estimates for the anomalous dimension for different values of
$N$, using both of our methods of approximation, are presented in
table \ref{table1}. We also plot, in figure \ref{graf_etas}, $N
\eta$ as a function of $1/N$. When $N$ is not too large, $N \simle
4$, strategy II introduces noticeable improvements. For example, for
$N = 1$, $\eta^*$ changes from $\eta^* \simeq 0.052$ using the first
approximation to $\eta^* \simeq 0.047$ using the improved one. These
values are to be compared with results obtained by other means. DE
gives $\eta^*=0$ at LO \cite{Morris94c} and $\eta^*=0.033$ at NNLO
\cite{Canet03}. As for the NLO, various results exist: using the
regulator of the present paper, one gets $\eta^*=0.050$
\cite{Canet02}, $\eta^*=0.054$ with a power-law regulator
\cite{Morris97} and $\eta^*=0.0467$ with an exponential regulator
\cite{Gersdorff00}; moreover, after an optimization procedure,
results move to $\eta^*=0.0470$, for a theta-like regulator, and
$\eta^*=0.0443$, for an exponential one \cite{Canet02}. In the
table, we present the best results with and without optimization.
Thus, even after the extra approximation introduced in the numerical
resolution strategy, the LO of the approximation procedure
introduced in \cite{BMW} yields a value of the anomalous dimension
comparable to that of the DE at NLO. All these NPRG values can be
compared with $\eta^* = 0.034(3)$ from the resummed 7 loop
calculation \cite{Guida98}, $\eta^*=0.0364(2)$ with high-T
calculation \cite{Campostrini02} and $\eta^*=0.0368(2)$ from MC
\cite{Deng03}.
\begin{center}
\begin{table}[ht]
\begin{scriptsize}
\begin{tabular}{|c|c|c|c|c|c|c|c|c|}
\hline
$N$ & $\eta$ I & $\eta$ II & $\eta$ DE-NLO & $\eta$ (best) & $\nu$ I & $\nu$ II & $\nu$ DE-NLO & $\nu$ (best) \\
\hline
0     & 0.0460   & 0.0385 & 0.039\cite{Gersdorff00} & 0.0284(25)\cite{Guida98} & 0.603   &   0.599 & 0.590\cite{Gersdorff00} & 0.5882(11) \cite{Guida98} \\
1   & 0.0519  & 0.0466  & 0.0467\cite{Gersdorff00} & 0.03639(15)\cite{Campostrini02} & 0.647 & 0.645 & 0.6307\cite{Gersdorff00} & 0.63012(16)\cite{Campostrini02} \\
    &         &    & 0.0443\cite{Canet02} & 0.0368(2)\cite{Deng03} &              &    & 0.6307\cite{Canet02} & 0.63020(12)\cite{Deng03} \\
2     & 0.0523  & 0.0474  & 0.049\cite{Gersdorff00} & 0.0381(2)\cite{Campostrini06} & 0.691 & 0.689 & 0.666\cite{Gersdorff00} & 0.6717(1) \cite{Campostrini06} \\
3     & 0.0496  & 0.0471  & 0.049\cite{Gersdorff00} & 0.0375(5)\cite{Campostrini01} & 0.729 &  0.728  & 0.704\cite{Gersdorff00} & 0.7112(5)\cite{Campostrini01}  \\
4     & 0.0455  & 0.0445  & 0.047\cite{Gersdorff00} & 0.0350(45)\cite{Guida98}  & 0.761  & 0.760  & 0.739\cite{Gersdorff00}  & 0.741(6)\cite{Guida98}   \\
      &         &     &    & 0.0365(10)\cite{Hasenbusch01}&     &    &     & 0.749(2)\cite{Hasenbusch01}\\
5     & 0.0409  & 0.0407 & & 0.031(3)\cite{Butti04} & 0.786  & 0.786  & & 0.764(4)\cite{Butti04}   \\
      &         &     &    & 0.034(1)\cite{Hasenbusch05} &   &      &        &  0.779(3)\cite{Hasenbusch05}\\
6     & 0.0368  &  &   & 0.029(3)\cite{Butti04} & 0.816  &    &      & 0.789(5)\cite{Butti04}  \\
7     & 0.0331  &  &   & 0.029 \cite{Antonenko98} & 0.838  &     &     & 0.811 \cite{Antonenko98}  \\
8     & 0.0298  &  &   & 0.027 \cite{Antonenko98}  & 0.856  &    &      & 0.830 \cite{Antonenko98} \\
9     & 0.0271  &  &  & 0.025 \cite{Antonenko98}  & 0.864  &     &     & 0.845 \cite{Antonenko98}   \\
10    & 0.0246  & 0.0253  & 0.028\cite{Gersdorff00} & 0.024 \cite{Antonenko98} & 0.882  & 0.882  & 0.881\cite{Gersdorff00}  & 0.859  \cite{Antonenko98}\\
20    & 0.0127  &  &     & 0.014 \cite{Antonenko98} & 0.941  &     &     & 0.930 \cite{Antonenko98}  \\
100   & 0.0025  & 0.00254 & 0.0030\cite{Gersdorff00} & 0.0027 \cite{Moshe03} & 0.990   & 0.990   & 0.990 \cite{Gersdorff00}   & 0.989\cite{Moshe03}     \\
large $N$ & $0.25/N$ & $0.25/N$ & & $0.270/N$ \cite{Moshe03} &$1 - 1.034/N$  & & & $1-1.081/N$ \cite{Moshe03} \\
\hline
\end{tabular}
\\
\end{scriptsize}
\caption{Critical exponents for the $O(N)$ model. We present our
results within the two calculation strategies, together with DE at
NLO and the best estimates, with their errors, when available. When
results of similar quality can be found in the literature, they are
both quoted (for discussion of DE results, see text).}
\label{table1}
\end{table}
\end{center}

As $N$ increases, our two calculation strategies differ less and
less, and eventually they give the same result. Most remarkably, our
yields are very close to those of the DE at NLO, at least for small
values of $N$. Note however that DE results can depend strongly on
the choice of the regulator: with a less reliable regulator, results
\cite{Morris97} depart from best estimates. Both this work and DE
results reach the correct large $N$ limit: $\eta = 0$. The large $N$
regime is better studied with the help of figure \ref{graf_etas}.
There, we present the best estimates, as well as DE and the present
results, together with the known analytical numbers for both the
large $N$ limit and its first order correction \cite{Moshe03}. One
can see that both our and DE calculations fail to reproduce the
analytical result (15\% error within DE, 8\% for this work). Most
remarkably, the large $N$ results seem to be better predicted with
both NPRG calculation than with the 6-loops resummed perturbation
calculation of ref. \cite{Antonenko98}; indeed, the latter clearly
fails to predict the large $N$ behavior. We shall be
back to the $1/N$ limit of our calculation in the last section of
the paper. Finally, it is interesting to notice that all results
present a peak in the value of $\eta$ around $N\sim 2 -3$ (see table
\ref{table1}).

Let us now turn to the critical exponent $\nu$. Close to
criticality, the effective renormalized mass at zero external field
(which is the same for the transversal and longitudinal propagating
modes in the symmetric phase), behaves as
\beq m_{ren}^2 (\rho = 0)= \frac{V'(\rho = 0)}{Z_\kappa} \simeq |r -
r_c|^{2\nu} \hskip 2 cm (\kappa \ll u) \eeq
where $r$ is the bare square mass and $r_c$ its value at
criticality. We have performed numerical calculations close to
criticality in order to determine the critical exponent $\nu$ for
various values of $N$. The results can be seen in table
\ref{table1}, together with those coming from DE at NLO and the best
estimates in the literature. Within both strategies of calculation
one gets almost the same results, for all values of $N$. As occuring
with the critical exponent $\eta$, here comparison of our results
with those from DE at NLO depends on the regulator used. For
example, for $N=1$, DE at NLO with the regulator of the present
paper gives $\nu=0.625$ \cite{Canet02}, $\nu=0.6307$ with the
exponential regulator \cite{Gersdorff00} and $\nu=0.618$ with a
power law regulator \cite{Morris97}. Optimization makes a tiny
difference \cite{Canet02}. In the present work, one gets $0.647$
(strategy I) and $0.645$ (strategy II). These numbers need to be
compared with 0.63012(16)\cite{Campostrini02} and
0.63020(12)\cite{Deng03}, from high-T and MC  calculations,
respectively.

\begin{figure}[t]
\begin{center}
\includegraphics*[scale=0.4,angle=-90]{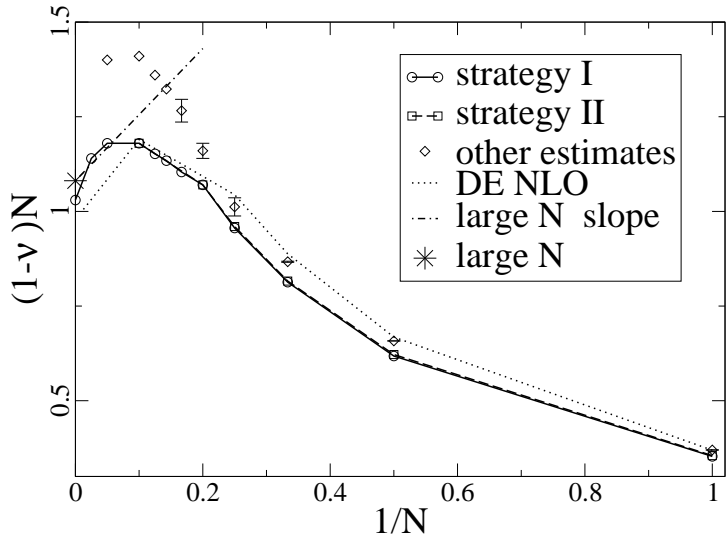}
\end{center}
\caption{ \label{nues} $N(1-\nu)$ as a function of $1/N$. This
quantity is useful to study the large $N$ behavior of the critical
exponent $\nu$}
\end{figure}

As for the large $N$ behavior, all calculations give the exact
limit, $\nu=1$. In order to strengthen the study of this regime, it
is convenient to plot the quantity $(1-\nu) N$, as we did in figure
\ref{nues}. Notice that these quantity is prone to have much larger
numerical error than $\nu$ as one moves to larger values of $N$. One
can see that both DE at NLO and the present calculation fails to
give the analytical large $N$ prediction for this quantity by not
more than a few percent. On the other side, as was the case for the
other critical exponent, 6-loops calculation results from ref.
\cite{Antonenko98} suffer from bigger errors when comparing with the exact $1/N$
expansion. We shall be back to these points in the next section.

Besides these already very competitive results, the present method
is fully devoted to calculate quantities that belong to a momentum
regime where neither direct perturbation theory nor scaling
properties can be used. These results are presented in the next
subsection.

\subsection{Ultraviolet and crossover regimes}

\begin{figure}[t]
\begin{center}
\includegraphics*[scale=0.4,angle=-90]{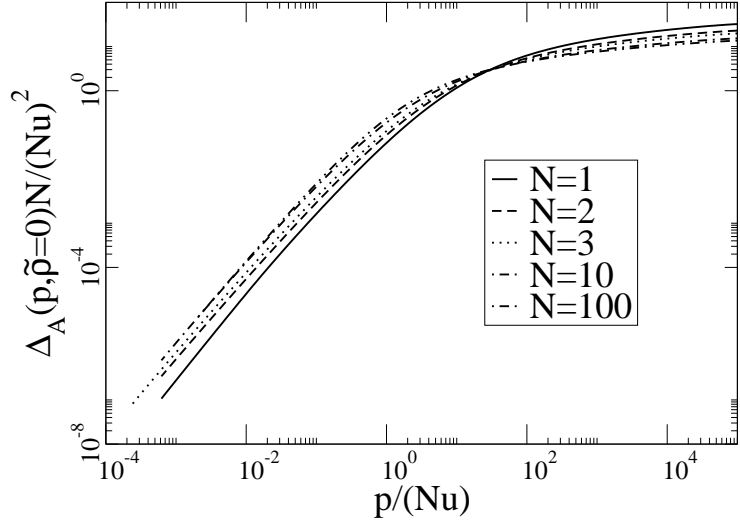}
\end{center}
\caption{ \label{dadpN2}  $\Delta_A(p; \rho = 0) N/(u/N)^2$, in $d=3$, for various $N$ at
criticality and zero external field, as a function of $p/u$. As soon as $N\simle 10$, the large $N$ limit is attained.}
\end{figure}

In figure \ref{dadpN2} the physical self energy at zero external
field, $\Delta_A(p,\tilde\rho=0,\kappa \to 0)$ is plotted for
various values of $N$. Notice that a simple dimensional analysis in
eq. (\ref{2pointclosed}) shows that the self-energy can be written
as $\Delta_A(p,\tilde\rho=0,\kappa=0)= u^2 \hat
\Delta_A(p/u,\tilde\rho=0,\kappa=0)$, where $\hat \Delta_A$ is a
dimensionless function of $p/u$. On the other hand, in the large $N$
limit the critical self-energy is of order $1/N$ (see e.g. the next
section). As this limit is taken so that $u N$ is kept constant, the
convenient function to plot is $ N \Delta_A(p/(N
u),\tilde\rho=0,\kappa=0) / (N u)^2$, which is shown in the figure.
One can see that, as soon as $N$ reaches $\sim 10$, corrections to
large $N$ behavior become small.

In the perturbative regime ($p\gg Nu$), one expects
$\Gamma^{(2)}_{ab}(p)\sim ((N+2)/3)
\delta_{ab}(u^2/96\pi^2)\log(p/u)$. In \cite{BMWnum} it has been
shown, for the $N = 1$ case, that the analytical solution of
eq.~(\ref{2pointclosed}) has this behavior, even though the
coefficient in front of the logarithm, $u^2/9\pi^4$, is
 8\% larger (this coefficient is a two-loop quantity and the LO of the approximation method introduced in
\cite{BMW} does not include all the 2-loop perturbative diagrams
exactly). The analysis of \cite{BMWnum} is generalizable for all $N$,
provided we include the multiplicative factor
$((N+2)/3)\delta_{ab}$. Our approximate numerical solutions, within
both methods, reproduce this result. As explained in \cite{BMW}, at
the NLO of the present approximation scheme, which is beyond the
scope of this paper, the contribution of the 2-loops diagrams is
completely included and the correct prefactor
$((N+2)u^2\delta_{ab}/288\pi^2)$ is recovered.

For purposes of probing the intermediate momentum region between the IR and the UV, we have
calculated the quantity
\begin{equation}\label{Delta_phi}
\frac{\delta
\langle\phi_a\phi_a\rangle}{u} = \frac{1}{u}\int \frac{d^3 p}{(2 \pi)^3} \left( \frac{1}{p^2 +
\Delta_A(p)} - \frac{1}{p^2} \right)
\end{equation}
which is very sensitive to the cross-over regime: the integrand in
eq. (\ref{Delta_phi}) is peaked at $p \sim (N u)/10$
\cite{Blaizot04}. In the $O(2)$ case and for $d=3$, this quantity
determines the shift of the critical temperature of the weakly
repulsive Bose gas \cite{Baym99}. It has thus been  much studied
recently using various methods, even for other values of $N$. In
particular, the large $N$ limit for this quantity has been
calculated analytically and found to be $1.05\times10^{-3}$
\cite{Baym00}. As a consequence, $\delta
\langle\phi_a\phi_a\rangle/u$ has been used as a benchmark for
non-perturbative approximations in the $O(N)$ model.

\begin{figure}[t]
\begin{center}
\includegraphics*[scale=0.4,angle=-90]{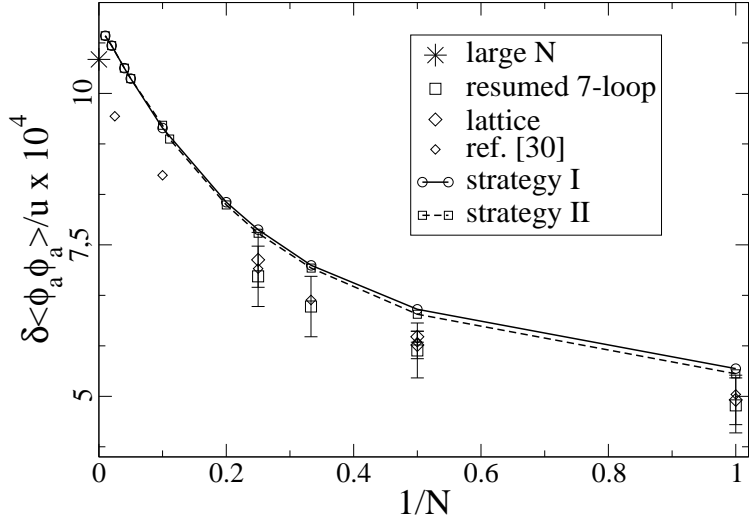}
\end{center}
\caption{ \label{graf_coefs} $\delta\langle\phi_a\phi_a\rangle/u$ as a function of $1/N$ }
\end{figure}

In this work, we have found the values for
$\delta\langle\phi_a\phi_a\rangle/u$ for some representative values
of $N$, using both of the methods of approximation for the potential
discussed in the previous section. The quantity we have chosen to
calculate, $\delta\langle\phi_a\phi_a\rangle/u$, is the one having a
finite large $N$ limit. The resulting curves are shown in figure
\ref{graf_coefs}, where they are compared with values obtained by
lattice calculations \cite{Arnold01,Kashurnikov01,Sun02}, resummed
7-loops calculations \cite{Kastening03,Kneur04} and to results
obtained in ref. \cite{Blaizot06}. Those numbers, with their
corresponding errors when available, are also presented in table
\ref{table2}. It can be seen that with our approximation strategy II
one gets slightly better results than with strategy I, but only for
small enough $N$. For all values of $N$ where lattice and/or 7-loops
resummed calculations exist, our results are almost within the error
bars of those calculations, except for $N=2$, where very precise
lattice results are available. In the large $N$ limit, one can see
that our results differ from the exact value by less than 4\%.
Please note that the large $N$ behavior of
$\delta\langle\phi_a\phi_a\rangle/u$ is given by $1/N$ corrections
to the self energy \cite{Baym00}, which are not calculated exactly
at this level of approximation. Notice also that, within the two
numerical strategies we use here, one gets the same large $N$ limit
for $\delta\langle\phi_a\phi_a\rangle/u$. Both in the table and the
figure we also present results from ref.~\cite{Blaizot06},
corresponding to an iterative ad-hoc calculation within NPRG. These
results correspond to an improved version of the NLO of that method,
which was a precursor of the procedure presented in \cite{BMW}. It
can be seen that, globally, these results are not significantly
better than those from the present paper, corresponding to an
approximate version of the method introduced in \cite{BMW}, at its
LO.

In summary, when trying to calculate quantities in the momenta
crossover region between the IR and the UV, the present
approximation scheme seems to be particularly well suited: already
with approximated versions of the LO of the method, one gets numbers
of about the same quality as those obtained with 7-loops resummed
perturbation theory or lattice calculations.

\begin{table}
\begin{tabular}{|c|c|c|c|c|c|}
\hline $N$ & lattice & 7-loops \cite{Kastening03} & ref.
\cite{Blaizot06} & strategy I & strategy II
\\ \hline
$1$ & $4.94 \pm 0.40 \times 10^{-4}$\cite{Sun02} & $4.85\pm 0.45
\times 10^{-4}$ & $5.03 \times 10^{-4}$ & $5.44 \times 10^{-4}$ &  $5.37 \times 10^{-4}$  \\
$2$ & $5.98 \pm 0.09 \times 10^{-4}$ \cite{Arnold01} & $ 5.76 \pm
0.45 \times 10^{-4}$ & $5.89 \times 10^{-4}$
& $6.44 \times 10^{-4}$ & $6.35 \times 10^{-4}$ \\
 & $5.85 \pm 0.22 \times 10^{-4}$\cite{Kashurnikov01} &   &   &   &    \\
$3$ &  & $6.48 \pm 0.50 \times 10^{-4}$ &
$6.57 \times 10^{-4}$ & $7.18 \times 10^{-4}$ & $7.12 \times 10^{-4}$  \\
$4$ & $7.25 \pm 0.45 \times 10^{-4}$\cite{Sun02} & $6.98 \pm
0.50$ & $7.12 \times 10^{-4}$ & $7.75 \times 10^{-4}$ & $7.69 \times 10^{-4}$  \\
$10$ &  &   & $8.66 \times 10^{-4}$ & $9.41 \times 10^{-4}$  &  $9.47 \times 10^{-4}$ \\
$100$ &  &   &    &  $1.10 \times 10^{-3}$  & $1.10 \times 10^{-3}$  \\
\hline
\end{tabular}
\\
\caption{Summary of available results for the universal quantity
$\delta\langle\phi_a\phi_a\rangle/u$. The analytically known exact
large $N$ limit for this quantity is equal to $1.05 \times
10^{-3}$\cite{Baym00}\label{table2}}
\end{table}

\section{Large-$N$ behavior}

\label{largeN}

As shown in \cite{BMW}, the leading order of the approximating
method introduced in that paper and used in the present one is exact
in the large $N$ limit, for any vertex function. This has been
verified numerically in the previous section, where the correct
large $N$ limits, $\eta=0$ and $\nu=1$, were extracted from the
$2$-point function. In this section, in order to go beyond these
results, we shall study the NLO of the $1/N$ expansion of the model,
to calculate both $\eta$ and $\nu$. The aim is to provide a test for
the approximation in a situation where it is not exact, as well as
to test the quality of the numerical calculations.

Before turning to this task, it is convenient to remind that the
large $N$ limit is taken so that $u N$ is constant and then
$\Gamma_\kappa \sim O (N) , V_\kappa \sim O (N)$, $\Gamma_A \sim
O(1)$,  $\Gamma_B \sim O(1/N)$  and $\rho \sim O (N)$
\cite{Moshe03}. In \cite{D'Attanasio97} a generic form of the
effective action in the large $N$ limit was found. Accordingly, a
general parametrization of the $2$-point function in this limit is
\beq \label{gamma2Ngrand}
\Gamma_{ab}^{(2)}(p,-p;\kappa;\phi)=(p^2+V'_\kappa
(\rho))\delta_{ab}+\phi_a\phi_b{\Gamma_B}(p,-p;\rho), \eeq where
$\Gamma_B$ is the function defined in eq.~(\ref{gamma2desc}). Using
this expression it is possible to show that, within this limit, the
flow equation for the potential at the leading order of DE (the
Local Potential Approximation, or LPA) is exact
\cite{D'Attanasio97}. It has also being exploited to solve
analytically the NPRG equations for any $n$-point function
\cite{BMW}. More precisely, doing the change of variables
\cite{Tetradis95} $(\kappa, \rho) \to (\kappa,W)$, where \beq
W=\frac{\partial V}{\partial \rho}, \eeq it is possible to find the
large $N$ limit for both the potential (or its inverse function $\rho = f(W)$) \cite{Tetradis95} and
$\Gamma_B$ \cite{BMW}:
\begin{align}\label{potLargeN}
f_\kappa(W) =&f_\Lambda(W)+ \frac{N}{2}\int \frac{d^dq}{(2\pi)^d}\left\{\frac{1}{q^2+W+R_\Lambda
 (q^2)}-\frac{1}{q^2+W+R_\kappa(q^2)} \right\}
 \\
\Gamma_B(p,-p;\rho) = &\frac{u}{3} \left( 1+\frac{Nu}{6}\int
\frac{d^dq}{(2\pi)^d}\frac{1}{q^2+W+R_\kappa
(q^2)}\frac{1}{(q+p)^2+W+R_\kappa ((q+p)^2)} \right)^{-1}
\label{gamma_b_largeN}.
\end{align} which in particular enables for an explicit solution for the $f_\kappa$ function (in $d=3$ and for the regulator (\ref{reg-litim})):
\begin{equation}\label{potLargeN2}
 \rho = f_\kappa(W) = \frac{3}{u} W - \frac{3}{u} r_R + \frac{3}{2} K_d N \kappa - K_d \frac{N}{2} \frac{\kappa^3}{\kappa^2 + W} + \frac{3}{2} N K_d W C(W)
\end{equation}
where $r_R=r_\Lambda+K_d N \Lambda u/3$ and $C(W)$ is defined by
\begin{equation}
C(W) \equiv \int_\kappa^\infty dq \frac{1}{q^2 +W}  =
\left\{
\begin{array}{ll}
\frac{1}{\sqrt{W}} \left( \frac{\pi}{2}-\mathrm{Arctan} \left(\frac{\kappa}{\sqrt{W}}\right) \right) \quad & (W>0) \\
\frac{1}{2\sqrt{-W}} \log \left( \frac{\sqrt{-W}+\kappa}{\sqrt{-W}
-\kappa} \notag \right) & (W<0)
\end{array}
\right.
\end{equation}
Observe that $W \sim  O (1)$.

\subsection{The critical exponent $\eta$}

The procedure to calculate the critical exponent $\eta$ is given by
eqs.~(\ref{ZkapparhoA})-(\ref{eta}) of appendix \ref{LPAp}. From the
definitions in eqs.~(\ref{ZkapparhoA}), (\ref{ZkapparhoB}) and
(\ref{adimchi}), one can see that $\chi_A \sim O(1)$ and $\chi_B
\sim O(1/N)$. As for the anomalous dimension, let us first observe
that at leading order, $\chi_A(\tilde \rho)\equiv 1$ (see
eq.~(\ref{gamma2Ngrand})). Accordingly, $\chi_A'(\tilde \rho)\sim
O(1/N^2)$. Thus, eq.~(\ref{eta}) implies that $\eta\sim O(1/N)$. In
order to find $\eta$, it is then necessary to get $\chi_B$ at its
leading order (i.e., $O(1/N)$) and $\chi_A$ at its next-to-leading
order (i.e., also $1/N$). The first function follows from
eq.~(\ref{ZkapparhoB}): expanding eq.~(\ref{gamma_b_largeN}) to
order $p^2$, one gets:
\begin{equation}\label{zetab}
Z_B = \frac{\frac{Nu^2}{18} K_3 \left[ \frac{1}{3}
\frac{\kappa^3}{(\kappa^2 +W)^3} + \frac{1}{4}
\frac{\kappa}{(\kappa^2 +W)^2} - \frac{1}{8W} \frac{\kappa}{\kappa^2
+W} + \frac{1}{8W} C(W) \right] }{\left[ 1 + \frac{Nu}{6} K_3 \left(
\frac{\kappa^3}{(\kappa^2 + W)^2} + \frac{3}{2}
\frac{\kappa}{\kappa^2 + W} + \frac{3}{2} C(W) \right) \right]^2}
\end{equation}
Finally, using eq.~(\ref{adimchi}) and introducing the dimensionless
quantities $\tilde{f} = f/(K_3 \kappa)$,  $w = W/\kappa^2 $ and
$\hat u= K_3 u/\kappa$, the exact leading order expression for
$\chi_B$ simply follows (remember that, for $N \to \infty$,
$Z_\kappa = 1$ and $\eta = 0$).

We now turn to the next-to-leading order expression for $\chi_A$. In order to do so, we shall solve its flow equation, i.e., eq.~(\ref{eqchia}), up to order $1/N$. One can then define:
\[
\chi_A = 1 + \frac{\hat{\chi}_A}{N}
\]
Introducing this expression in the flow equation (\ref{eqchia}) and keeping only leading terms in $1/N$, one gets:
\begin{multline}\label{chia}
\partial_t \hat{\chi}_A -(d-2) \tilde{\rho} \hat{\chi}_A' + \frac{N\hat{\chi}_A'}{(1+w)^2} \\
= \eta N + \frac{8 N \tilde{\rho}}{1+w+2\tilde{\rho}w'}
\frac{\chi_Bw'}{(1+w)^2}
 - \frac{2N \chi_B}{(1+w)^2} - \frac{4N\tilde{\rho}w'^2}{(1+w+2\tilde{\rho}w')^2(1+w)^2}
\end{multline}
Notice that, by construction, $\hat{\chi}_A \sim O(1)$ (and then
$\hat{\chi}_A' \sim O(1/N)$). Accordingly, in the right-hand-side of
eq. (\ref{chia}) only the leading order expressions of both $\chi_B$
and $w$ are needed. They follow from eqs. (\ref{potLargeN2}) and
(\ref{zetab}), after going to dimensionless variables.

Going to variables $(\kappa,w)$, using $\partial_t \vert_\rho =
\partial_t \vert_w + \partial_\kappa w \partial_w$ and
eq.~(\ref{flow_of_w}), we end up with (where we define $\partial_w
\tilde f_\kappa(w) \equiv \tilde f_\kappa'(w)$)
\begin{multline}\label{chia_largeN}
 \partial_t \hat{\chi}_A(w) - 2w \partial_w \hat{\chi}_A(w) \\
= \eta N + \frac{8 N \tilde{f}(w)}{1+w+2\tilde{f}(w)/\tilde{f}'}
\frac{\chi_B/\tilde{f}'(w)}{(1+w)^2} - \frac{2N \chi_B(w)}{(1+w)^2}
-\frac{4N\tilde{f}/(\tilde{f}'(w))^2}{(1+w+2\tilde{f}(w)/\tilde{f}'(w))^2(1+w)^2}
\end{multline}
Notice that it is convenient to work in the particular case $w=0$,
i.e., at the minimum of the potential, where the equation is much
simpler. Going to the IR fixed point, $\kappa \partial_\kappa
\chi_A(w) \equiv 0$ (and, equivalently, $\hat{u} \to \infty$) it
then follows that:
\begin{equation}\label{etalargeN}
\eta^* = \frac{1}{4N} + O\left(\frac{1}{N^2}\right)
\end{equation}

One expects the calculation of $\eta^*$ to be independent of the
renormalization point. However, once approximations are done there
is normally a small dependence on the precise renormalization point.
In this case, though, observe that all over this calculation we need
not to consider the specific renormalization condition introduced in
eqs.~(\ref{def-Zk}) and (\ref{defetak}) (i.e., $\rho_0 = 0$). It
then turns out that the obtained value of $N\eta^*$, when $N \to
\infty$, does not depend on the chosen renormalization point.

The analytical value of $\eta^*$ obtained above is reproduced by our
numerical results (see figure \ref{graf_etas} and table 1). The
exact value for the $1/N$ correction to $\eta^*$ in $O(N)$ models is
known  \cite{Zinn-Justin02}: $8/(3 \pi^2) \simeq 0.27$. It turns out
that the error of our (both numerical and analytical) calculation
turns out to be of about $8 \%$, which is smaller than the error
involved in both DE and 6 loops resummed perturbative calculations
(see figure \ref{graf_etas}).

A surprising fact is that the ratio between our result ($1/4$) and
the correct number ($8/(3 \pi^2)$) is exactly the same as the ratio
between the correct result for the ultraviolet coefficient in front
of the logarithm, $(N+2)u^2/(288\pi^2)$, and ours, $(N+2)
u^2/(27\pi^4)$. It thus seems that, at least for $\eta$, the LO of
the approximate procedure introduced in \cite{BMW} misses the second
order of the $1/N$ expansion by the same amount that it misses the
second order of the perturbative expansion.

It is important to notice that the calculation presented above is
valid for both of the methods of approximation described in section
3 of this paper. This is due to the fact that in this section we
only needed the large $N$ limit of the potential, which is already
exact in the case of the approximation used in strategy I.

\subsection{The critical exponent $\nu$}

Let us now move on to the analysis of the $1/N$ correction for the
behavior of the critical exponent $\nu$. We are dealing here with a
critical point for a second order phase transition, with only one IR
unstable direction under the RG flow. The exponent $\nu$ is related
to the eigenvalue of the linearized flow in this relevant direction.

It will be useful to start with a study of the large $N$ limit for
this quantity. Due to the significant simplification brought by
working close to $w=0$ in the study of $\eta$ just presented, a
similar strategy will now be adopted, defining

\[
w = a (\tilde{\rho} - \rho_0) + \frac{b}{2} (\tilde{\rho} -
\rho_0)^2 + O((\tilde{\rho} - \rho_0)^3)
\]
with
\begin{align}\notag
w(\rho_0) & = 0 & a & = w'(\rho_0) & b & = w''(\rho_0)
\end{align}

It is easy to find equations for this set of variables, which in
the large $N$ limit decouple \cite{D'Attanasio97}, yielding the
equations
\begin{align}
\partial_t \rho_0 & = -(d-2) \rho_0 + N \label{rho0eq_LO}\\
\partial_t a & = (d-4) a + 2Na^2 \label{aeq_LO} \\
\partial_t b & = (2d-6) b +6N(ab - a^3)
\end{align}
It is then possible to find the non-Gaussian fixed point value for
these quantities in the large $N$ limit:
\begin{equation}\label{rho0*_LO}
\rho_0^* = \frac{N}{d-2}
\end{equation}
\begin{equation}\label{a*_LO}
a^* = \frac{4-d}{2N}
\end{equation}
\begin{equation}\label{b*_LO}
b^* = \frac{6Na^{*3}}{2(d-3) + 6Na^*} = \frac{3}{4N^2}
\frac{(4-d)^3}{6-d}
\end{equation}

The flow equations for $\rho_0(t)$ and $a(t)$, eqs (\ref{rho0eq_LO})
and (\ref{aeq_LO}), can be solved analytically, yielding
\begin{equation}\label{rho0_LO}
\rho_0(t) = \rho_0^* + P e^{-(d-2)t}
\end{equation}
\begin{equation}\label{a_LO}
a(t) = a^* \frac{1}{A e^{(4-d)t} + 1}
\end{equation}

These expressions, together with straightforward calculations for
the behavior of $b(t)$ or any other higher order coupling near the
fixed point, clearly show that, in the large $N$ limit, the only
unstable quantity under the RG flow, as $t\to-\infty$ ($\kappa \to
0$), is $\rho_0(t)$. This allows for the determination of the
expected large $N$ critical exponent $\nu^{LO} = 1/(d-2)$; in
particular $\nu = 1$ for $d=3$.

Moving on to the analysis of the NLO in the $1/N$ expansion, we will
from now on specialize to the simpler case of the first resolution
strategy of section \ref{strategies}. Observe that results should
not differ by much in the case of the improved strategy of section
\ref{improved}, as the numerical calculations for both
approximations seem to converge rapidly as $N$ increases.

The NLO flow equation for $\rho_0$, for the case of strategy I is

\begin{equation}\label{rho0eq_NLO}
\partial_t \rho_0 = - (d-2+\eta) \rho_0 + \left(1 -
\frac{\eta}{d+2} \right) \bigg[ (N-1) + \frac{3+ 2\rho_0
\frac{b}{a}}{(1+2\rho_0a)^2} \bigg]
\end{equation}

Notice here that the functions $a(t)$ and $b(t)$ need only be
known in their large $N$ limit. We will focus in solutions for this
flow equations that are near the fixed point. We can then write, for
the quantities in the r.h.s. of eq. (\ref{rho0eq_NLO})
\begin{align}
\rho_0 (t) & = \rho_0^* + \epsilon \hat{\rho_0}e^{-t(d-2)} \\
b(t) & = a^* + \epsilon \hat{a}(t) \\
b(t) & = b^* + \epsilon \hat{b}(t) \\
\eta(t) & = \eta^* + \epsilon \hat{\eta} \hat{\rho_0} e^{-t(d-2)}
\label{etahat_def}
\end{align}
where for $\rho_0$ and $\hat{\eta}$ we are explicitly taking out the
known large $N$ $t$ dependence, and in the case of $\hat{\eta}$ we
are defining it with an ansatz to be justified later. As for the
l.h.s. of eq. (\ref{rho0eq_NLO}) we can in principle also write
$\rho_0 (t) = \rho_0^* + \epsilon \hat{\rho_0}^{(1)}(t) e^{-t(d-2)}$
Expanding eq. (\ref{rho0eq_NLO}) to order $\epsilon$ one can reach
the expression (throwing away terms of order (1/N))
\begin{multline}
\partial_t \hat{\rho}_0^{(1)}(t) = - \eta^* \hat{\rho}_0 - \hat{\eta} \rho_0^* \hat{\rho}_0 -
\frac{\hat{\eta}N}{d+2} \hat{\rho}_0 + \bigg[ \frac{2}{a^*}
\big(\rho_0^* \hat{b}e^{t(d-2)} + \hat{\rho}_0 b^* - \rho_0^* b^*
\frac{\hat{a}e^{t(d-2)}}{a^*} \big) \label{rho0_hat}\\ -
\frac{4\big(3 + \frac{2}{a^*} \rho_0^* b^* \big) \big(\rho_0^*
\hat{a}e^{(t(d-2))} + \hat{\rho}_0 a^* \big)}{1 + 2 \rho_0^*
a^*}\bigg] \frac{1}{(1 + 2 \rho_0^* a^*)^2}
\end{multline}

As already stated, we know that in the large $N$ limit $\hat{a}(t)$
and $\hat{b}(t)$ go to zero after $t$ gets negative enough.
Nonetheless, we cannot in principle assert the same for the behavior
of $\eta$, and in fact we will show below that, as $\eta$ is already
a quantity of order $O(1/N)$, the IR unstable direction of the flow
has components in the linearized directions of both $\hat{\rho}_0$
and $\hat{\eta}$. If we consider $|t|$ large enough for $\hat{a}$
and $\hat{b}$ to be negligible, the following flow equation is found
\begin{equation}
\frac{1}{\hat{\rho}_0} \partial_t \hat{\rho}_0^{(1)}(t) = - \eta^* -
\hat{\eta} \rho_0^* - \frac{\hat{\eta}N}{d+2} + \bigg[
\frac{2b^*}{a^*} - \frac{12a^* +8 \rho_0^* b^*}{1 + 2 \rho_0^*
a^*}\bigg] \frac{1}{(1 + 2 \rho_0^* a^*)^2}
\end{equation}

Knowing the large $N$ behavior for $\hat{\rho}_0^{(1)}(t)$ (see eq.
(\ref{rho0_hat})), one can assume that, at NLO of the $1/N$
expansion
\[
\hat{\rho}_0^{(1)}(t) = \hat{\rho}_0 e^{\left(- t \frac{y_1}{N}
\right)} \simeq \hat{\rho}_0 \left(1 - t \frac{y_1}{N} + \ldots
\right)
\]
where, in $d=3$, it is easy to check that $y_1 = - \nu^{(1)}$, i.e.,
the correction for $\nu$ at NLO in $1/N$ ($\nu = 1/(d-2) +
\nu^{(1)}/N$). The equation for $\hat{\rho}_0^{(1)}$ can then be
written as an equation for $\nu^{(1)}$
\begin{equation}
\frac{\nu^{(1)}}{(d-2)^2N} = - \eta^* - \hat{\eta} \rho_0^* -
\frac{\hat{\eta}N}{d+2} - \bigg[\frac{2b^*}{a^*} - \frac{12a^* +8
\rho_0^* b^*}{1 + 2 \rho_0^* a^*}\bigg] \frac{1}{(1 + 2 \rho_0^*
a^*)^2}
\end{equation}
expression which, when evaluated in $d=3$, using eqs.
(\ref{rho0eq_LO}), (\ref{a*_LO}) and (\ref{b*_LO}) and also
(\ref{etalargeN}) yields the relationship
\begin{equation}
 \nu^{(1)} = -1 - \frac{6N\hat{\eta}}{5}
\end{equation}
the $-1$ in the last expression is the bulk of the required result,
as $\hat{\eta}$ is expected to bring only a small contribution, which will be calculated in what follows.

Notice before, that the calculation so far presented is independent
of the renormalization scheme, just as for the case of $\eta$ of the
last subsection. In what follows we will calculate the correction
$\hat{\eta}$, which does depend on the point where the value of
$\eta$ is fixed. Nevertheless, as stated in the last section, the
numerical results found for the quantity $(1-\nu)N$ suffer from a
quite large numerical error for large values of $N$. This makes
impossible to distinguish the results between the two strategies
presented in this paper, or between different renormalization schemes.

The starting point for the calculation of $\hat \eta$ is eq. (\ref{chia_largeN}) for
$\hat{\chi}_A(\tilde{\rho},t)$. In it, scheme dependence can only
come from the running of $\eta_\kappa$ and from boundary conditions.
When one considers the fixed point solution for $\hat{\chi}_A$ all
the scheme dependence should come from the boundary conditions, as
we have already shown that $\eta^*$, the fixed point solution for
$\eta_\kappa$, is scheme independent. Therefore, the fixed point
equation is of the form $w\partial_w  \hat{\chi}_A = N\eta^* +
F^*(w)$, where the r.h.s is scheme independent and thus the only
consequence of the renormalization scheme on the fixed point solution is an additive
constant.

Near the fixed point, in the r.h.s. $F(w,t) = F^*(w) + \epsilon
e^{-t(d-2)}\hat{F}(w)$, and similarly $\hat{\chi}_A(w,t) =
\hat{\chi}_A^*(w) + \epsilon e^{-t(d-2)}\tilde{\chi}_A$. At order
$\epsilon$
\begin{equation}\label{chia_tilde}
(d-2) \tilde\chi_A(w)+2w \partial_w\tilde \chi_A(w)+N\hat \eta=-\tilde F(w)
\end{equation}

Defining the variable $\bar{\chi}_A \equiv \tilde{\chi}_A +
N\hat{\eta}/(d-2)$, it is straightforward to find the only solution
to the eq. (\ref{chia_tilde}) which is a regular function for all
values of $w$, and also turns out to be scheme independent.
\begin{equation}\label{chia_techo}
\bar\chi_A(w)= -\frac{1}{2}|w|^{\frac{2-d}{2}}\int_0^w
dw'|w'|^{\frac{d-4}{2}}\tilde F(w')
\end{equation}
As for the function $\tilde F(w)$ it is easily obtained by an order
$\epsilon$ expansion of the function $F(w)$, taking into account
that $(f')^* = (f^*)'$, and the fact that $\chi_B$ has a contractive
IR behavior.
\begin{align}
\tilde F(w)&=\frac{-\hat \rho_0 N}{(1+w)^2} \Bigg\{\frac{8
\chi_B^*(w)/(\tilde {f}^*)'(w)} {1+w+2 \tilde
f^*(w)/(\tilde{f}^*)'(w)} -\frac{16 \tilde f^*
\chi_B^*(w)/((\tilde{f}^*)'(w))^2}
{(1+w+2 \tilde f^*(w)/(\tilde{f}^*)'(w))^2} \notag \\
&-\frac{4}{((f^*)'(w))^2}
\frac{1}{(1+w+2 \tilde f^*(w)/(\tilde{f}^*)'(w))^2} \notag\\
&+\frac{16}{((f^*)'(w))^3}\frac{\tilde{f}^*(w)} {(1+w+2 \tilde
f^*(w)/(\tilde{f}^*)'(w))^3} \Bigg\}
\end{align}
where the factor $\hat\rho_0$ comes from the $t$ dependence of the
function $\tilde{f}(w,t) = \tilde{f}^*(w) + \epsilon \hat{\rho_0}
e^{t(d-2)}$. Notice that this factor justifies our initial ansatz for
$\hat \eta$, eq (\ref{etahat_def}).

After calculating the integral in eq. (\ref{chia_techo}), which can
only be done numerically, one can impose the chosen renormalization
prescription and obtain the numerical value of $\hat{\eta}$ which
would complete the calculation of $\nu^{(1)}$. If we take the
renormalization condition $\tilde \chi_A(w(\rho=0))=0$ used for the
numerical results already presented, we find
\begin{equation}
\hat \eta=\frac{d-2}{N}\bar \chi_A(w^*(\rho=0))
\end{equation}

In $d=3$, numerical integration yields $\bar{\chi}_A(w^*(\rho=0) \simeq
0.028238$, which allows for finding the NLO $1/N$ expansion
correction for $\nu$ under our approximation scheme:
\begin{equation}\label{nu}
\nu \simeq 1 - \frac{1.034}{N}
\end{equation}
which is to be compared with the exact NLO $1/N$ expansion result
\cite{Moshe03} $\nu = 1 - 1.081/N$; in this case, then, the
analytical error induced by the approximation used in this work is
of the order of $5\%$, that is, less than that involved in the determination
of $\eta$. Our numerical results seem to reproduce the analytical
value given by eq. (\ref{nu}), even though the numerical uncertainty
for the quantity $N(1-\nu)$ grows for $N \to \infty$ (see fig.
\ref{nues}).

As in the case of $\eta$, the results of this work seem to better
reproduce the large $N$ behavior of the model, when compared with DE
or 6-loops resummed perturbative calculations. In fact, as can be seen in fig.
\ref{nues}, there appears to be a notorious disagreement between the
expected large $N$ behavior and the resummed perturbative results of
ref. \cite{Antonenko98}.

\section{Conclusions and Perspectives}

In this paper we calculate the 2-point functions of the $O(N)$
model, in all ranges of momenta, i.e., either the IR, the UV and the
crossover intermediate region. We use a method proposed in
\cite{BMW}, which allows for an approximate solution for any
$n$-point function NPRG flow equation. Although this method can be
systematically improved, in the present work we consider only the
leading order (LO) of this approximation procedure. In fact, in
order to deal with a simpler numerical problem, on top of the
already approximated flow equation we have done  extra assumptions
to simplify propagators. Moreover, we have tested two strategies:
while the first one is simpler (we call it strategy I), it misses
the $2$-loop exactness of the potential; within the second one
(strategy II), which implies a small extra numerical effort, the
$2$-loop expression is recovered. We have shown that both strategies
can be justified; in addition, they are exact both perturbativelly
and in the large $N$ limit.

We have calculated various quantities to gauge the quality of the
2-point function obtained within both strategies. In the IR regime,
our solutions correctly reproduce the scaling properties. We have
calculated two critical exponents, $\eta$ and $\nu$, for some
representative values of $N$. We got numbers in reasonable agreement
with the best known results available in the literature, obtained
either with lattice or 7-loops resummed perturbative calculations.
Within strategy II, our numbers turn out to be of the same quality
of those obtained with Derivative Expansion (DE) at NLO. Notice
however that DE is only well suited to reproduce these scaling
quantities. When going to large values of $N$, our numerical results
seem to reproduce the analytically known large $N$ limits, exactly
at the LO and approximately at the NLO in the $1/N$ expansion. This
also happens with DE results but not with those from a 6-loops
resummed perturbative calculation \cite{Antonenko98} (in fact, as
soon as $N \simge 5 - 10$, results from ref. \cite{Antonenko98} seem
to deviate from the correct large $N$ behavior).

The UV behavior of the $O(N)$ model 2-point function is of course
very well known: it presents a logarithmic shape, the pre-coefficent
following from a 2-loop calculation. One can analytically prove that
the self-energy following from both our strategies does have this
logarithmic shape. Nevertheless, as the LO of the approximate method
of \cite{BMW} does not exactly include all 2-loop diagrams
contributing to the $2$-point function, the pre-coefficent is missed
by 8\%. We have checked that our numerical results, within both
methods, reproduce these analytical predictions.

The intermediate crossover region between IR and UV regimes is
certainly the most challenging one. In order to check the quality of
our $2$-point function, we have calculated a quantity which is
particularly sensitive to this momentum regime. Of course, this
number cannot be obtained with DE and only lattice or resummed
perturbative calculations, as well as analytical large $N$ results,
exist. In this case, our method is largely competitive with the best
available results, except for the $N=2$ case, where very precise
lattice results are known. We miss the analytical large $N$ limit
for this quantity by 4\%; notice that the calculated quantity is of
NLO in a $1/N$ expansion, which is not completely included in our
approximate calculation.

Let us finally add a remark concerning our results: as expected,
strategy II is more accurate than strategy I, but the differences
get very small when going to large values of $N$. This is related to
the fact that in the large $N$ limit the approximate propagators we
used in strategy I become exact.

To our knowledge, this is the first time that the 2-point function
of the $O(N)$ model is calculated with such overall acceptable
properties. Within the NPRG, two ad-hoc calculations can be quoted
here. First, in refs. \cite{Ledowski04} an approximate solution of
the flow equation including only a limited number of vertices could
be calculated, but the result is unstable when trying to improve the
method \cite{kopietz}. On the other side, in refs
\cite{Blaizot04,Blaizot05,Blaizot06}, an ad-hoc method which can be
considered as the precursor of the method presented in \cite{BMW} is
used to get a 2-point function with all the expected properties;
nevertheless, acceptable numbers only follow after an improved NLO
order variation of the method, and after a lengthy analytical and
numerical effort.

We have also studied analytically the $1/N$ expansion of the
critical exponents $\eta$ and $\nu$. In fact, as the first order in
the expansion ($\eta=0$ and $\nu=1$) is trivially reproduced by our
calculation we went up to the second order. In the case of $\eta$,
we found a result 8\% smaller than the correct one.  As for $\nu$,
we get $(1-\nu) N = 1.03 + O(1/N^2)$, while the correct result is
$(1-\nu) N = 1.081 + O(1/N^2)$. Here, we miss the correct result by
only 5\%. Finally, let us notice that our results seem to reproduce
the expected analytical predictions. This is a strong support for
our numerics.

To conclude, we observe that this approximate method to calculate
2-point functions at finite momenta within the NPRG seems to work
properly for all values of $N$. It reproduces, already at the LO,
all the expected behavior of the self-energy, giving, in most of the
regions of $N$ and $p$, results similar to those of the best
accepted values in the literature. Moreover, in the large $N$ limit,
one gets the best numbers, if comparing with DE or resummed 6-loop
calculations. The result of this paper thus extend and confirm those
already obtained in \cite{BMWnum} for the Ising universality class.

For the near future, two works are to be developed. As a natural
extension, it would be interesting to go beyond strategies I and II
and try to numerically solve the full approximate equation for the
2-point function, eq.~(\ref{2pointclosed}). On the other side, all
these works call for an application of this method to
non-perturbative problems, as e.g. QCD, where only approximate
procedures including a limited number of vertices have been
considered up to now.

\appendix

\section{The $p=0$ sector}

\label{LPAp}

Within strategy I, one first solves the flow equations of the
``$p=0$'' sector, i.e., those for the potential $V_\kappa(\rho)$ and
the field renormalization constant $Z_\kappa$ (see section
\ref{strategies}). In fact, this is nothing but a variation of the
Local Potential Approximation (LPA), which includes a ($\rho$
independent) field renormalization constant \cite{Berges00}.
Although this is a very well known procedure, its yields depend on
the precise definition one uses for $Z_\kappa$. As discussed in the
main text, in order to describe correctly the scaling regime, in our
case the flow equation for $Z_\kappa$ has to be compatible with the
approximate eq.~(\ref{2pointclosed}) for the 2-point function.

Let us first consider the flow equation for the potential. It
follows from eq.~(\ref{eqforV}) with the $I^{(1)}$ functions given by
the analytic expression from eq. (\ref{In-anal}). In fact, it is
numerically preferable to solve the flow equation for
$w_\kappa(\tilde \rho) = \partial_{\tilde \rho}
v_\kappa(\tilde\rho)$ which reads: \beq\label{flow_of_w}
\partial_t w_\kappa\!=\! -(2\!-\!\eta_\kappa) w_\kappa + (d -
2+\!\eta_\kappa) \tilde\rho w'_\kappa
 -\! \left( 1\!-\!\frac{\eta_\kappa}{d+2} \right) \!\left( \frac{(N\!-\!1) w'_\kappa}{(1+w_\kappa)^2} \!+\!
 \frac{3w'_\kappa + 2\tilde\rho w''_\kappa}{(1+w_\kappa+2 \tilde\rho w'_\kappa)^2} \right).\nonumber\\
\eeq

As for $Z_\kappa$, let us first define:
\begin{align}
Z_A(\rho,\kappa)\equiv & 1+\left.\frac{\partial \Delta_A(p;\rho)}{\partial
p^2}\right|_{p=0},  \label{ZkapparhoA} \\
Z_B(\rho,\kappa)\equiv & \left.\frac{\partial \Delta_B(p;\rho)}{\partial
p^2}\right|_{p=0} \label{ZkapparhoB}
\end{align}
whose flow equations follow from those of $\Delta_A$ and $\Delta_B$. Then, one  imposes the definition of $Z_\kappa$ given by eq. (\ref{def-Zk}). In fact, as the functions $Z_A(\rho,\kappa)$ and $Z_B(\rho,\kappa)$, as well as the potential, are zero momentum quantities, it is preferable to define dimensionless quantities (see eq. (\ref{adim})):
\begin{align} \label{adimchi}
\chi_A(\tilde \rho)\equiv &\frac{Z_A(\rho)}{Z_\kappa}, &
\chi_B(\tilde \rho)\equiv &\frac{Z_B(\rho)}{\kappa^{4-d}Z_\kappa^2
K_d^{-1}}.
\end{align}

The flow equations for $\chi_A$ and $\chi_B$ are then:
\begin{align}\label{eqchia}
\partial_t \chi_A & =  \eta \chi_A + (d - 2 + \eta) \tilde{\rho} \chi_A' + 2 \left(1 - \frac{\eta}{d+2} \right) \nonumber \\
&  \Bigg[ \frac{2\tilde{\rho}}{(1 + w + 2\tilde{\rho}w')^2}
\frac{1}{1+w} (2 \chi_A' w') +
  \frac{2\tilde{\rho}}{(1 + w)^2} \frac{2 \chi_B w'}{1 + w + 2\tilde{\rho}w'} \nonumber \\
&- \frac{1}{2} \frac{1}{(1 + w + 2\tilde{\rho}w')^2} (\chi_A' +
2\tilde{\rho} \chi_A'') -
  \frac{1}{2} \frac{1}{(1 + w)^2} ((N-1) \chi_A' + 2 \chi_B) \Bigg] \nonumber \\
&- \frac{2\tilde{\rho}}{(1 + w + 2\tilde{\rho}w')^2} \frac{1}{(1 +
w)^2} 2w'^2
\end{align}
\begin{align}\label{eqchib}
\partial_t \chi_B & =  (d - 2 + 2\eta)\chi_B + (d - 2 + \eta) \tilde{\rho} \chi_B' + 2 \left(1 - \frac{\eta}{d+2} \right) \nonumber \\
&  \Bigg[ \frac{2(N-1)}{(1+w)^3} \chi_B w' +
\frac{1}{(1+w+2\tilde{\rho}w')^3}
  \Big\{6 \chi_A' w' + 12 \chi_B w' \nonumber \\
&+ 4\tilde{\rho}(\chi_A' w'' + 3\chi_B' w' + 2 \chi_B w'') + 8 \tilde{\rho}^2 \chi_B' w''\Big\} \nonumber \\
&- \frac{1}{(1+w+2\tilde{\rho}w')^2} \frac{2 \chi_A' w'}{(1+w)} -
    \frac{1}{(1+w)^2} \frac{2 \chi_B w'}{(1+w+2\tilde{\rho}w')} \nonumber \\
&- \frac{1}{2} \frac{(N-1)}{(1+w)^2} \chi_B' - \frac{1}{2}
    \frac{1}{(1+w+2\tilde{\rho}w')^2} (5 \chi_B' + 2\tilde{\rho}\chi_B'') \nonumber \\
&+ \left(\frac{1}{1+w+2\tilde{\rho}w'} + \frac{1}{1+w}\right)
\frac{1}{1+w}
    \frac{\chi_B w'}{1+w+2\tilde{\rho}w'} \Bigg] \nonumber \\
&- (N-1) \frac{w'^2}{(1+w)^4} - \frac{1}{(1+w+2\tilde{\rho}w')^4}(9
w'^2 +
12\tilde{\rho}w'w'' \nonumber \\
&+ 4\tilde{\rho}^2w''^2) + \frac{1}{(1+w)^2}
\frac{2w'^2}{(1+w+2\tilde{\rho}w')^2}
\end{align}

Finally, the value of $\eta_\kappa$ follows from eq. (\ref{defetak}):
\begin{equation}\label{eta}
\eta_\kappa =  \frac{(N \chi_A'(\tilde{\rho} = 0) + 2
\chi_B(\tilde{\rho} = 0))}{(1 + w(\tilde{\rho} = 0))^2 + \frac{(N
\chi_A'(\tilde{\rho} = 0) + 2 \chi_B(\tilde{\rho} = 0))}{d+2}} .
\end{equation}

In strategy I, the $p=0$ sector of the theory thus follows from the simultaneous solution of the 3 flow equations (\ref{flow_of_w}), (\ref{eqchia}) and (\ref{eqchib}), together with eq.~(\ref{eta}).
Eq.~(\ref{flow_of_w}) is solved starting from the initial condition
at $\kappa=\Lambda$:
\beq\label{wLambdadez}
w_\kappa(\tilde\rho,\kappa=\Lambda) = \hat m_\Lambda^2 + \hat
g_\Lambda \tilde\rho,
\eeq
where the dimensionless parameters $ \hat m_\Lambda$ and $ \hat
g_\Lambda$ are related to the parameters $r$ and $u$ of the
classical action, eq. (\ref{Sclassical}), by ($d=3$)
\beq\label{relclas}
 \hat m_\Lambda^2 =\frac{r}{\Lambda^2},\qquad \hat g_\Lambda = \frac{u }{\Lambda}\frac{K_3}{3},
\eeq
As for the initial conditions for the functions $\chi$, they follow from the definitions in eqs. (\ref{ZkapparhoA}),    (\ref{ZkapparhoB}), (\ref{defdelta_a_2}) and (\ref{defdelta_b_2}), together with eq. (\ref{Sclassical}):
\beq
\chi_A(\tilde\rho;\kappa=\Lambda)=1 \hskip 0.3cm , \hskip 0.5cm  \chi_B(\tilde\rho;\kappa=\Lambda)=0.
\eeq

The parameter $r$ is adjusted in order to be at criticality at zero external field, i.e., in order to have a vanishing physical mass. This is achieved imposing the dimensionless mass $m^2_\kappa(\tilde\rho=0)=w_\kappa(\tilde\rho=0)$ to reach a finite value when $\kappa \ll u$.

\section{The function $J$}
\label{functionf}

In this section we present the expressions for the function
$J_d^{(3)}(p;\rho;\kappa)$, defined in eq.~(\ref{defJ}). Using the
approximate propagators from eqs.~(\ref{propGT}), (\ref{propGL}),
(\ref{propGTp+q}) and (\ref{propGLp+q})   the integral can be done
analytically (in $d=3$). One gets (see ref. \cite{BMWnum} for the
$N=1$ case):

\noindent a) $\bar p>2, \hat m_\beta^2<0 $.
\begin{align}\label{Jlarge}
&\hspace{-.2cm}J_{3,\alpha \beta}^{(3)}(p;\kappa;\tilde\rho)
=\frac{1}{\kappa Z_\kappa^2(2\pi)^2}\frac{1}{(1+\hat m_\alpha^2)^2}\left\lbrace 2
+\frac{\eta}{2}\left(-\frac{5}{3}+\bar p^2-3\hat m_\beta^2\right) \right.
\nonumber \\
&+\frac{1}{2\bar p}\left[-1+\frac{\eta}{4}+\left(\bar p+\sqrt{-\hat m_\beta^2}\right)^2
\left(1-\frac{\eta}{2}
+\frac{\eta}{4}\left(\bar p+\sqrt{-\hat m_\beta^2}\right)^2\right)\right]
\log\left(\frac{\bar p-1+\sqrt{-\hat m_\beta^2}}{\bar p+1+\sqrt{-\hat m_\beta^2}}
\right) \nonumber \\
&+\left.\frac{1}{2\bar p}\left[-1+\frac{\eta}{4}+\left(\bar p-\sqrt{-\hat m_\beta^2}
\right)^2\left(1-\frac{\eta}{2}
+\frac{\eta}{4}\left(\bar p-\sqrt{-\hat m_\beta^2}\right)^2\right)\right]
\log\left(\frac{\bar p-1-\sqrt{-\hat m_\beta^2}}{\bar p+1-\sqrt{-\hat m_\beta^2}}
\right) \right\rbrace \nonumber \\
&=\frac{1}{\kappa Z_\kappa^2(2\pi)^2}\frac{2}{(1+\hat m_\alpha^2)^2}\left\lbrace
\frac{4}{\bar p^2}\left(\frac{1}{3}-\frac{\eta}{15}\right)+\frac{4}{\bar p^4}\left(
\frac{1}{15}-\frac{\eta}{105}
-\frac{\hat m_\beta^2}{3}+\frac{\eta\hat m_\beta^2}{15}\right)+
\mathcal{O}(1/(\bar p^6)) \right\rbrace . \nonumber \\
\end{align}
b) $\bar p\le 2, \hat m_\beta^2<0$.
\begin{align}\label{Jsmall}
&\hspace{-.2cm}J_{3, \alpha \beta}^{(3)}(p;\kappa;\tilde\rho)
=
\frac{1}{\kappa Z_\kappa^2(2\pi)^2}\frac{1}{(1+\hat m_\alpha^2)^2}\left\lbrace
-1+\frac{\eta}{4}+\frac{\eta\hat m
_\beta^2}{4}+\bar p\left(\frac{3}{2}
-\frac{\eta}{8}-\frac{7\eta\hat m_\beta^2}{8}\right)
-\frac{3\eta}{4}\bar p^2\right. \nonumber \\
&+\frac{25\eta}{48}\bar p^3+\frac{1}{1+\hat m_\beta^2}\left(\frac{4}{3}
-\frac{4\eta}{15}-\bar p+\frac{\eta}{3}\bar p^2
+\left(\frac{1}{12}-\frac{\eta}{6}\right)\bar p^3+\frac{\eta}{120}\bar p^5\right)
\nonumber \\
&+\hspace{-.2cm}
\frac{1}{2\bar p}\left[1-\frac{\eta}{4}
-\left(\bar p+\sqrt{-\hat m_\beta^2}\right)^2
\left(1-\frac{\eta}{2}+\frac{\eta}{4}
\left(\bar p+\sqrt{-\hat m_\beta^2}\right)^2\right)\right]
\log \left(\frac{\bar p+1+\sqrt{-\hat m_\beta^2}}{1+\sqrt{-\hat m_\beta^2}}\right)
\nonumber \\
&+\left.\hspace{-.2cm}\frac{1}{2\bar p}\left[1-\frac{\eta}{4}-
\left(\bar p-\sqrt{-\hat m_\beta^2}\right)^2\left(1-\frac{\eta}{2}+\frac{\eta}{4}
\left(\bar p-\sqrt{-\hat m_\beta^2}\right)^2\right)\right]
\log\left(\frac{\bar p+1-\sqrt{-\hat m_\beta^2}}{1-\sqrt{-\hat m_\beta^2}}\right)
\right\rbrace\nonumber \\
&=
\frac{1}{\kappa Z_\kappa^2(2\pi)^2}\frac{1}{(1+\hat m_\alpha^2)^2}\left\lbrace
\frac{4}{3(1+\hat m_\beta^2)}\left(1-\frac{\eta}{5}\right)-\frac{2}{3(1+\hat m_\beta^2)^2}\bar p^2 \right.\nonumber \\
&+\left. \frac{2+\eta-2\hat m_\beta^2+\eta \hat m_\beta^2}{6(1+\hat m_\beta^2)^3}\bar p^3
-\frac{2(1+\eta-5\hat m_\beta^2+\eta \hat m_\beta^2)}{15(1+\hat m_\beta^2)^4}\bar p^4+\mathcal{O}(\bar p^5)\right\rbrace.
\end{align}
c) $\bar p>2, m_\beta^2\ge 0$.
\begin{align}\label{Jlarge2}
&\hspace{-.2cm}J_{3, \alpha \beta}^{(3)}(p;\kappa;\tilde\rho) =\frac{1}{\kappa
Z_\kappa^2(2\pi)^2}\frac{1}{(1+\hat m_\alpha^2)^2}\left\lbrace 2
+\frac{\eta}{2}\left(-\frac{5}{3}+\bar p^2-3\hat m_\beta^2\right) \right. \nonumber \\
&+\frac{1}{\bar p}\left[\left(-1+\frac{\eta}{4}+(\bar p^2-\hat m_\kappa^2)\left(1-\frac{\eta}{2}
+\frac{\eta}{4}(\bar p^2-\hat m_\beta^2)\right)-\eta\hat m_\beta^2\bar p^2\right)\frac{1}{2}\log\left(\frac{(\bar p-1)^2+\hat m_\kappa^2}{(\bar p+1)^2+\hat m_\kappa^2}
\right) \right.\nonumber\\
&\left.\left.-2\hat m_\beta \bar p \left(1-\frac{\eta}{2}+\frac{\eta}{2}(\bar p^2-\hat m_\beta^2)\right)\left(
\mathrm{Arctan} \left(\frac{\hat m_\beta}{\bar p-1}\right)-\mathrm{Arctan} \left(\frac{\hat m_\beta}{\bar p+1}\right)\right)\right]\right\}\nonumber \\
&=\frac{1}{\kappa Z_\kappa^2(2\pi)^2}\frac{1}{(1+\hat
m_\alpha^2)^2}\left\lbrace
\frac{4}{\bar p^2}\left(\frac{1}{3}-\frac{\eta}{15}\right)\right. \nonumber\\
&\left.+\frac{1}{105\,\bar p^4}\left(7-35\hat m_\beta^2+\eta(-1+7\hat m_\beta^2)\right)+\mathcal{O}(1/(\bar p^6)) \right\rbrace \nonumber \\
\end{align}
d) $\bar p\leq 2, m_\beta^2\ge 0$.
\begin{align}\label{largeJ4}
&\hspace{-.2cm}J_{3, \alpha \beta}^{(3)}(p;\kappa;\tilde\rho) =\frac{1}{\kappa
Z_\kappa^2(2\pi)^2(1+\hat m_\alpha^2)^2}\left\lbrace
-1+\frac{\eta}{4}+\frac{\eta \hat m_\beta^2}{4}+\bar
p\left(\frac{3}{2}-\frac{\eta}{8}-\frac{7\eta \hat
m_\beta^2}{8}\right)
-\frac{3\eta \bar p^2}{4} \right. \nonumber \\
&+ \frac{25\eta \bar p^3}{48}+\frac{1}{1+\hat m_\beta^2}\left(\frac{4}{3}-\frac{4\eta}{15}-\bar p+\frac{\eta \bar p^2}{3}
+\frac{\bar p^3}{12}-\frac{\eta \bar p^3}{6}+\frac{\eta \bar p^5}{120} \right) \nonumber \\
&+\frac{1}{\bar p}\left[\left(1-\frac{\eta}{4}-(\bar p^2-\hat m_\beta^2)\left(1-\frac{\eta}{2}+\frac{\eta}{4}
(\bar p^2-\hat m_\beta^2)\right)+\eta \hat m_\beta^2 \bar p^2\right)\frac{1}{2}\log \left(\frac{(\bar p+1)^2+\hat m_\beta^2}{1+\hat m_\beta^2}\right) \right.\nonumber \\
&+\left.\left.2\hat m_\beta \bar p \left(1-\frac{\eta}{2}+\frac{\eta}{2}(\bar p^2-\hat m_\beta^2)\right)
\left(\mathrm{Arctan} \left(\frac{\hat m_\beta}{\bar p+1}\right)-\mathrm{Arctan} \left(\hat m_\beta\right)\right)\right]\right\} \nonumber \\
&=\frac{1}{\kappa Z_\kappa^2(2\pi)^2(1+\hat
m_\alpha^2)^2}\left\lbrace
\frac{4}{3(1+\hat m_\beta^2)}\left(1-\frac{\eta}{5}\right)-\frac{2}{3(1+\hat m_\beta^2)^2}\bar p^2 \right.\nonumber \\
&+\left. \frac{2+\eta-2\hat m_\beta^2+\eta \hat m_\beta^2}{6(1+\hat m_\beta^2)^3}\bar p^3
-\frac{2(1+\eta-5\hat m_\beta^2+\eta \hat m_\beta^2)}{15(1+\hat m_\beta^2)^4}\bar p^4+\mathcal{O}(\bar p^5)\right\rbrace .
\end{align}

Finally, going to dimensionless variables one can define:

\beq
\tilde{J}_{\alpha \beta}(\tilde{p}; \tilde{\rho}) \equiv J_{\alpha \beta}(p; \rho)  \frac{\kappa Z_\kappa^2}{K_3}
\eeq

\bibliographystyle{unsrt}

\end{document}